\documentstyle[12pt]{article}

\skewchar\fivmi='177
\skewchar\sixmi='177
\skewchar\sevmi='177
\skewchar\egtmi='177
\skewchar\ninmi='177
\skewchar\tenmi='177
\skewchar\elvmi='177
\skewchar\twlmi='177
\skewchar\frtnmi='177
\skewchar\svtnmi='177
\skewchar\twtymi='177
\def\@magscale#1{ scaled \magstep #1}
\font\twfvmi  = ammi10   \@magscale5 
\skewchar\twfvmi='177
\skewchar\fivsy='60
\skewchar\sixsy='60
\skewchar\sevsy='60
\skewchar\egtsy='60
\skewchar\ninsy='60
\skewchar\tensy='60
\skewchar\elvsy='60
\skewchar\twlsy='60
\skewchar\frtnsy='60
\skewchar\svtnsy='60
\skewchar\twtysy='60
\font\twfvsy  = amsy10   \@magscale5 
\skewchar\twfvsy='60

\catcode`@=11
\def\un#1{\relax\ifmmode\@@underline#1\else
        $\@@underline{\hbox{#1}}$\relax\fi}
\catcode`@=12

\let\du=\d                      
\let\um=\H                      

\def\a{\alpha}
\def\b{\beta}

\def\d{\delta}
\def\e{\epsilon}
\def\f{\phi}
\def\g{\gamma}

\def\j{\psi}
\def\k{\kappa}
\def\l{\lambda}
\def\m{\mu}
\def\n{\nu}

\def\r{\rho}
\def\s{\sigma}

\font\sc=font005                        
\def\Sc#1{{\hbox{\sc #1}}}      
\font\ooo=circle10                      
\font\ro=manfnt                         
\def\kcl{{\hbox{\ro 6}}}                
\def\kcr{{\hbox{\ro 7}}}                
\def\ktl{{\hbox{\ro \char'134}}}        
\def\ktr{{\hbox{\ro \char'135}}}        
\def\kbl{{\hbox{\ro \char'136}}}        
\def\kbr{{\hbox{\ro \char'137}}}        

\def\ip{{=\!\!\! \mid}}                                    
\def\bo{{\raise.15ex\hbox{\large$\Box$}}}               
\def\pr{\prod}                                          
\def\TH{{\raise.2ex\hbox{$\displaystyle \bigodot$}\mskip-4.7mu \llap H \;}}
\def\face{{\raise.2ex\hbox{$\displaystyle \bigodot$}\mskip-2.2mu \llap {$\ddot
        \smile$}}}                                      

\def\sp#1{{}^{#1}}                              
\def\Tilde#1{{\widetilde{#1}}\hskip 0.03in}                     
\def\Hat#1{\widehat{#1}}                        
\def\Bar#1{\overline{#1}}                       
\def\leftrightarrowfill{$\mathsurround=0pt \mathord\leftarrow \mkern-6mu
        \cleaders\hbox{$\mkern-2mu \mathord- \mkern-2mu$}\hfill
        \mkern-6mu \mathord\rightarrow$}
\def\dvec#1{\vbox{\ialign{##\crcr
        \leftrightarrowfill\crcr\noalign{\kern-1pt\nointerlineskip}
        $\hfil\displaystyle{#1}\hfil$\crcr}}}           
\def\dt#1{{\buildrel {\hbox{\LARGE .}} \over {#1}}}     

\def\frac#1#2{{\textstyle{#1\over\vphantom2\smash{\raise.20ex
        \hbox{$\scriptstyle{#2}$}}}}}                   
\def\ha{\frac12}                                        
\def\sfrac#1#2{{\vphantom1\smash{\lower.5ex\hbox{\small$#1$}}\over
        \vphantom1\smash{\raise.4ex\hbox{\small$#2$}}}} 
\def\bfrac#1#2{{\vphantom1\smash{\lower.5ex\hbox{$#1$}}\over
        \vphantom1\smash{\raise.3ex\hbox{$#2$}}}}       
\def\afrac#1#2{{\vphantom1\smash{\lower.5ex\hbox{$#1$}}\over#2}}    

\newskip\humongous \humongous=0pt plus 1000pt minus 1000pt
\def\caja{\mathsurround=0pt}
\def\eqalign#1{\,\vcenter{\openup2\jot \caja
        \ialign{\strut \hfil$\displaystyle{##}$&$
        \displaystyle{{}##}$\hfil\crcr#1\crcr}}\,}
\newif\ifdtup
\def\panorama{\global\dtuptrue \openup2\jot \caja
        \everycr{\noalign{\ifdtup \global\dtupfalse
        \vskip-\lineskiplimit \vskip\normallineskiplimit
        \else \penalty\interdisplaylinepenalty \fi}}}
\def\li#1{\panorama \tabskip=\humongous                         
        \halign to\displaywidth{\hfil$\displaystyle{##}$
        \tabskip=0pt&$\displaystyle{{}##}$\hfil
        \tabskip=\humongous&\llap{$##$}\tabskip=0pt
        \crcr#1\crcr}}

\def\ref#1{$\sp{#1)}$}

\parskip=\medskipamount                 
\thicklines                         

\def\oldheadpic{                                
        \setlength{\unitlength}{.4mm}
        \thinlines
        \par
        \begin{picture}(349,16)
        \put(325,16){\line(1,0){4}}
        \put(330,16){\line(1,0){4}}
        \put(340,16){\line(1,0){4}}
        \put(335,0){\line(1,0){4}}
        \put(340,0){\line(1,0){4}}
        \put(345,0){\line(1,0){4}}
        \put(329,0){\line(0,1){16}}
        \put(330,0){\line(0,1){16}}
        \put(339,0){\line(0,1){16}}
        \put(340,0){\line(0,1){16}}
        \put(344,0){\line(0,1){16}}
        \put(345,0){\line(0,1){16}}
        \put(329,16){\oval(8,32)[bl]}
        \put(330,16){\oval(8,32)[br]}
        \put(339,0){\oval(8,32)[tl]}
        \put(345,0){\oval(8,32)[tr]}
        \end{picture}
        \par
        \thicklines
        \vskip.2in}
\def\oldtitle#1#2#3#4{\oldheadpic\begin{center}\vglue.5in{\large\bf #1}\\[.6in]
        {#2}\\[.1in] {\it Department of Physics and Astronomy}\\
        {\it University of Maryland, College Park, MD 20742}\\[.6in]
        Physics Publication \#{#3}\\ {#4}\\[1.5in] {\bf Abstract}\\[.1in]
        \end{center} \begin{quotation}}                 
\def\oldTitle#1#2#3#4#5#6#7{\oldheadpic\begin{center} \vglue .4in
        {\large\bf #1}\\[.4in]
        {#2}\\[.1in] {\it Department of Physics and Astronomy}\\
        {\it University of Maryland, College Park, MD 20742}\\[.1in]
        {#3}\\[.1in] {\it {#4}}\\ {\it {#5}}\\[.4in]
        Physics Publication \#{#6}\\ {#7}\\[.5in] {\bf Abstract}\\[.1in]
        \end{center} \begin{quotation}}                 
\def\border{                                            
        \setlength{\unitlength}{1mm}
        \newcount\xco
        \newcount\yco
        \xco=-24
        \yco=12
        \begin{picture}(140,0)
        \put(\xco,\yco){$\ktl$}
        \advance\yco by-1
        {\loop
        \put(\xco,\yco){$\kcl$}
        \advance\yco by-2
        \ifnum\yco>-240
        \repeat
        \put(\xco,\yco){$\kbl$}}
        \xco=158
        \yco=12
        \put(\xco,\yco){$\ktr$}
        \advance\yco by-1
        {\loop
        \put(\xco,\yco){$\kcr$}
        \advance\yco by-2
        \ifnum\yco>-240
        \repeat
        \put(\xco,\yco){$\kbr$}}
        \put(-20,11){\tiny University of Maryland Elementary Particle
Physics University of Maryland Elementary Particle Physics University of
Maryland Elementary Particle Physics}
        \put(-20,-241.5){\tiny University of Maryland Elementary
Particle Physics University of Maryland Elementary Particle Physics
University of Maryland Elementary Particle Physics}
        \end{picture}
        \par\vskip-8mm}
\def\bordero{                                           
        \setlength{\unitlength}{1mm}
        \newcount\xco
        \newcount\yco
        \xco=-24
        \yco=12
        \begin{picture}(140,0)
        \put(\xco,\yco){$\ktl$}
        \advance\yco by-1
        {\loop
        \put(\xco,\yco){$\kcl$}
        \advance\yco by-2
        \ifnum\yco>-240
        \repeat
        \put(\xco,\yco){$\kbl$}}
        \xco=158
        \yco=12
        \put(\xco,\yco){$\ktr$}
        \advance\yco by-1
        {\loop
        \put(\xco,\yco){$\kcr$}
        \advance\yco by-2
        \ifnum\yco>-240
        \repeat
        \put(\xco,\yco){$\kbr$}}
        \put(-20,12){\ooo
bacdefghidfghghdhededbihdgdfdfhhdheidhdhebaaahjhhdahba
hgdedge
   hgfdiehhgdigicba}
        \put(-20,-241.5){\ooo
ababaighefdbfghgeahgdfgafagihdidihiidhiagfedhadbfd
ecdcdfa
   gdcbhaddhbgfchbgfdacfediacbabab}
        \end{picture}
        \par\vskip-8mm}
\def\headpic{                                           
        \indent
        \setlength{\unitlength}{.4mm}
        \thinlines
        \par
        \begin{picture}(29,16)
        \put(165,16){\line(1,0){4}}
        \put(170,16){\line(1,0){4}}
        \put(180,16){\line(1,0){4}}
        \put(175,0){\line(1,0){4}}
        \put(180,0){\line(1,0){4}}
        \put(185,0){\line(1,0){4}}
        \put(169,0){\line(0,1){16}}
        \put(170,0){\line(0,1){16}}
        \put(179,0){\line(0,1){16}}
        \put(180,0){\line(0,1){16}}
        \put(184,0){\line(0,1){16}}
        \put(185,0){\line(0,1){16}}
        \put(169,16){\oval(8,32)[bl]}
        \put(170,16){\oval(8,32)[br]}
        \put(179,0){\oval(8,32)[tl]}
        \put(185,0){\oval(8,32)[tr]}
        \end{picture}
        \par\vskip-6.5mm
        \thicklines}
\def\title#1#2#3#4{\border\headpic {\hbox to\hsize{#4 \hfill UMDEPP #3}}\par
        \begin{center} \vglue .5in {\large\bf #1}\\[.6in]
        {#2}\\[.1in] {\it Department of Physics and Astronomy}\\
        {\it University of Maryland, College Park, MD 20742}\\[1.5in]
        {\bf Abstract}\\[.1in] \end{center} \begin{quotation}}  
\def\Title#1#2#3#4#5#6#7{\border\headpic
        {\hbox to\hsize{#7 \hfill UMDEPP #6}}\par
        \begin{center} \vglue .4in {\large\bf #1}\\[.4in]
        {#2}\\[.1in] {\it Department of Physics and Astronomy}\\
        {\it University of Maryland, College Park, MD 20742}\\[.1in]
        {#3}\\[.1in] {\it {#4}}\\ {\it {#5}}\\[.5in] {\bf Abstract}\\[.1in]
        \end{center} \begin{quotation}}                 
\def\endtitle{\end{quotation}\newpage}                  

\def\sect#1{\bigskip\medskip \goodbreak \noindent{\bf {#1}} \nobreak \medskip}
\def\refs{\sect{References} \footnotesize \frenchspacing \parskip=0pt}
\def\Item{\par\hang\textindent}

\begin{document}

\def\sqrtrf{{\sqrt{1-\fracmm1{r^4}}}}
\def\sinhth{\sinh\vartheta}
\def\coshth{\cosh\vartheta}
\def\sinhthsq{\sinh^2\vartheta}
\def\coshthsq{\cosh^2\vartheta}

\def\gg{{\hbox{\sc g}}}
\def\nt{$~N=2$~}
\def\gg{{\hbox{\sc g}}}
\def\nt{$~N=2$~}
\def\tr{{\rm tr}}
\def\Tr{{\rm Tr}}
\def\mpl#1#2#3{Mod.~Phys.~Lett.~{\bf A{#1}} (19{#2}) #3}
\def\prep#1#2#3{Phys.~Rep.~{\bf #1} (19{#2}) #3}

\def\scst{\scriptstyle}
\def\itrema{$\ddot{\scriptstyle 1}$}
\def\Bo{\bo{\hskip 0.03in}}
\def\lrad#1{ \left( A {\buildrel\leftrightarrow\over D}_{#1} B\right) }

\def\ula{{\underline a}} \def\ulb{{\underline b}} \def\ulc{{\underline c}}
\def\uld{{\underline d}} \def\ule{{\underline e}} \def\ulf{{\underline f}}
\def\ulg{{\underline g}} \def\ulh{{\underline h}} \def\ulm{{\underline m}}
\def\uln{{\underline n}}
\def\ul#1{\underline{#1}}
\def\ulp{{\underline p}} \def\ulq{{\underline q}} \def\ulr{{\underline r}}
\def\uls{{\underline s}}

\def\plpl{{+\!\!\!\!\!{\hskip 0.009in}{\raise -1.0pt\hbox{$_+$}}
{\hskip 0.0008in}}}
\def\mimi{{-\!\!\!\!\!{\hskip 0.009in}{\raise -1.0pt\hbox{$_-$}}
{\hskip 0.0008in}}}

\def\items#1{\\ \item{[#1]}}
\def\ul{\underline}
\def\un{\underline}
\def\-{{\hskip 1.5pt}\hbox{-}}

\def\kd#1#2{\d\du{#1}{#2}}

\def\fracmm#1#2{{{#1}\over{#2}}}
\def\footnotew#1{\footnote{\hsize=6.5in {#1}}}
\def\low#1{{\raise -3pt\hbox{${\hskip 1.0pt}\!_{#1}$}}}

\def\ip{{=\!\!\! \mid}}
\def\ze{\zeta^{+}}
\def\zeb{{\bar \zeta}^{+}}
\def\umb{{\underline {\bar m}}}
\def\unb{{\underline {\bar n}}}
\def\upb{{\underline {\bar p}}}
\def\um{{\underline m}}
\def\up{{\underline p}}
\def\Phib{{\Bar \Phi}}
\def\Phit{{\tilde \Phi}}
\def\Phibt{{\tilde {\Bar \Phi}}}
\def\Db{{\Bar D}_{+}}
\def\gg{{\hbox{\sc g}}}
\def\nt{$~N=2$~}

\border\headpic {\hbox to\hsize{October 1992 \hfill UMDEPP 93--79}}\par
\begin{center}
\vglue .16in

{\large\bf Self--Dual Supergravity and Supersymmetric Yang--Mills \\
Coupled to Green--Schwarz Superstring$\,$}\footnote{This work is
supported in part by NSF grant \# PHY-91-19746.} \\[.1in]

\baselineskip 10pt
\vskip 0.22in

Hitoshi ~NISHINO
{}~~~ \\[.2in]

{\it Department of Physics} \\ [.015in]
{\it University of Maryland at College Park}\\ [.015in]
{\it College Park, MD 20742-4111, USA} \\[.1in]
and\\[.1in]
{\it Department of Physics and Astronomy} \\[.015in]
{\it Howard University} \\[.015in]
{\it Washington, D.C. 20059, USA} \\[.12in]

\vskip 0.32in

{\bf Abstract}\\[.1in]
\end{center}

\begin{quotation}

{\baselineskip 5pt We present the {\it canonical} set of superspace
constraints for self-dual supergravity, a ``self-dual'' tensor
multiplet and a self-dual Yang-Mills multiplet with
$~N=1~$ supersymmetry in the space-time with signature $(+,+,-,-)$.
For this set of constraints, the consistency of the self-duality
conditions on these multiplets with supersymmetry is manifest.
The energy-momentum tensors of all the self-dual ``matter'' multiplets vanish,
to be consistent with the self-duality of the Riemann tensor.
In particular, the special significance
of the ``self-dual'' tensor multiplet is noted.  This result fills the
gap left over in our previous series of papers, with respect to the consistent
couplings among the self-dual matter multiplets.  We also couple these
non-trivial backgrounds to a Green-Schwarz superstring $~\s\-$model,
under the requirement
of invariance under fermionic (kappa) symmetry.  The finiteness of the
self-dual supergravity is discussed, based on its ``off-shell''
structure.  A set of exact
solutions for the ``self-dual'' tensor and self-dual
Yang-Mills multiplets for the gauge group $~SL(2)$~ on
self-dual gravitational instanton background is given, and its
consistency with the Green-Schwarz string ~$\s\-$model is demonstrated.
}

\endtitle

\def\doit#1#2{\ifcase#1\or#2\fi}
\def\[{\lfloor{\hskip 0.35pt}\!\!\!\lceil}
\def\]{\rfloor{\hskip 0.35pt}\!\!\!\rceil}
\def\delsl{{{\partial\!\!\! /}}}
\def\caldsl{{\calD\!\!\! /}}
\def\calO{{\cal O}}
\def\asym{({\scriptstyle 1\leftrightarrow \scriptstyle 2})}
\def\Lag{{\cal L}}
\def\du#1#2{_{#1}{}^{#2}}
\def\ud#1#2{^{#1}{}_{#2}}
\def\dud#1#2#3{_{#1}{}^{#2}{}_{#3}}
\def\udu#1#2#3{^{#1}{}_{#2}{}^{#3}}
\def\calD{{\cal D}}
\def\calM{{\cal M}}
\def\tildef{{\tilde f}}
\def\calDsl{{\calD\!\!\!\! /}}

\def\Hat#1{{#1}{\large\raise-0.02pt\hbox{$\!\hskip0.038in\!\!\!\hat{~}$}}}
\def\hati{{\hat{I}}}
\def\dt{$~D=10$~}
\def\alp{\alpha{\hskip 0.007in}'}
\def\oalp#1{\alp^{\hskip 0.007in {#1}}}
\def\naive{{{na${\scriptstyle 1}\!{\dot{}}\!{\dot{}}\,\,$ve}}}
\def\items#1{\vskip 0.05in\Item{[{#1}]}}
\def\item#1{\Item{#1}}

\def\pl#1#2#3{Phys.~Lett.~{\bf {#1}B} (19{#2}) #3}
\def\np#1#2#3{Nucl.~Phys.~{\bf B{#1}} (19{#2}) #3}
\def\prl#1#2#3{Phys.~Rev.~Lett.~{\bf #1} (19{#2}) #3}
\def\pr#1#2#3{Phys.~Rev.~{\bf D{#1}} (19{#2}) #3}
\def\cqg#1#2#3{Class.~and Quant.~Gr.~{\bf {#1}} (19{#2}) #3}
\def\cmp#1#2#3{Comm.~Math.~Phys.~{\bf {#1}} (19{#2}) #3}
\def\jmp#1#2#3{Jour.~Math.~Phys.~{\bf {#1}} (19{#2}) #3}
\def\ap#1#2#3{Ann.~of Phys.~{\bf {#1}} (19{#2}) #3}
\def\prep#1#2#3{Phys.~Rep.~{\bf {#1}C} (19{#2}) #3}
\def\ptp#1#2#3{Prog.~Theor.~Phys.~{\bf {#1}} (19{#2}) #3}
\def\ijmp#1#2#3{Int.~Jour.~Mod.~Phys.~{\bf {#1}} (19{#2}) #3}
\def\nc#1#2#3{Nuovo Cim.~{\bf {#1}} (19{#2}) #3}
\def\ibid#1#2#3{{\it ibid.}~{\bf {#1}} (19{#2}) #3}

\def\szet{{${\scriptstyle \b}$}}
\def\ula{{\un a}}
\def\ulb{{\un b}}
\def\ulc{{\un c}}
\def\uld{{\un d}}
\def\ulA{{\un A}}
\def\ulM{{\underline M}}
\def\cdm{{\Sc D}_{--}}
\def\cdp{{\Sc D}_{++}}
\def\vTheta{\check\Theta}
\def\Pisl{{\Pi\!\!\!\! /}}

\def\fracmm#1#2{{{#1}\over{#2}}}
\def\gg{{\hbox{\sc g}}}
\def\half{{\fracm12}}
\def\ha{\half}

\def\frac#1#2{{\textstyle{#1\over\vphantom2\smash{\raise -.20ex
        \hbox{$\scriptstyle{#2}$}}}}}                   

\def\fracm#1#2{\hbox{\large{${\frac{{#1}}{{#2}}}$}}}

\def\Dot#1{\buildrel{_{_{\hskip 0.01in}\bullet}}\over{#1}}
\def\dt#1{\Dot{#1}}
\def\uln{{\underline n}}
\def\Tilde#1{{\widetilde{#1}}\hskip 0.015in}
\def\Hat#1{\widehat{#1}}

\def\Dot#1{\buildrel{_{_\bullet}}\over{#1}}
\def\dt#1{\Dot{#1}}
\def\Hat#1{\widehat{#1}}
\def\Bar#1{\overline{#1}}

\oddsidemargin=0.03in
\evensidemargin=0.01in
\hsize=6.5in
\textwidth=6.5in

\centerline{\bf I.~~Introduction}
\medskip\medskip

        We have recently presented a series of papers [1-5] on the subject
of self-dual (supersymmetric) gauge theories in the space-time with the
signature $~(+,+,-,-)$.\footnotew{We call this $~D=(2,2)$~ space-time.
When the signature is not important, we also use the expression
$~D=4$.}~~The importance of these theories stems
from the conjecture [6] that {\it all} the (bosonic) exactly soluble models in
lower dimensions are obtained by some kinds of dimensional
reductions of self-dual Yang-Mills (SDYM) theory [7],
originally {\it without} supersymmetry.  Based on this, we have
developed a generalized conjectured that
possibly {\it all} the exactly soluble {\it supersymmetric}
models in lower-dimensions are to
be from self-dual {\it supersymmetric} YM (SDSYM) theory [1-5].
We have also shown that {\it extended} supersymmetries are compatible
with these self-dual gauge theories [4,5].  The
purely bosonic SDYM and self-dual gravity (SDG) fields can couple to $~N=2$~
superstring [8], which is supposed to be finite to all string loops.
To have manifest target space-time supersymmetry,
we have also presented the Green-Schwarz (GS) superstring
[3] on SDSYM and self-dual supergravity
(SDSG) backgrounds, in which the target
space backgrounds have manifest $~N=1$~ supersymmetry.

        Motivated by these developments, we have shown in our recent paper [9]
that $~D=2$~ soluble systems such as $~N=2$~ supersymmetric KdV system and
$~N=1$~ super Toda theory can be actually embedded into the $~D=1,\,N=2$~
SDSYM.  We have further presented some
exact solutions for the coupled system of SDSG, SDSYM and what we call
``self-dual'' tensor multiplet (SDTM) on a gravitational instanton
background [10].

        In the superspace formulation in our previous paper [3],
we have given what is called
beta-function-favoured constraints (BFFC) [11,12] for the $~N=1$~ supergravity
(SG), tensor multiplet (TM) and Yang-Mills multiplet (SYM), and shown
the consistency of their couplings to a GS string $~\s\-$model.
We have also shown the
satisfaction of the condition of vanishing $~\b\-$functions, such that
these backgrounds are consistent with conformal invariance of the
GS superstring.  In particular, the $~\b\-$function calculation had been
drastically simplified due to the BFFC we have chosen.
These BFFC, however, had a drawback, because some additional field
redefinitions were needed to maintain the manifest consistency between
the self-duality (SD) conditions and supersymmetry [3].
For example, due to the non-vanishing supertorsion
$~T\du{\ula\ulb} \ulc\neq 0$,
we could not show the validity of the SD condition
$~R\du{\ula\ulb}{\ulc\uld} = (1/2) \e\du{\ula\ulb}{\ule\ulf}
R\du{\ule\ulf}{\ulc\uld}$, without performing an additional
super-Weyl rescaling [3].  This is because the BFFC system was {\it not} in
the {\it canonical} frame with the manifest SD conditions [3].

        According to a more recent analysis in Ref.~[13], the closed
$~N=2$~ superstring in the Neveu-Schwarz-Ramond formulation will have
the $~N=8$~ SDSG background as its consistent background.  Even though this
$~N=8$~ multiplet is {\it irreducible}, it can be truncated into lower
$~N~(1\le N \le 4)$~ multiplets consistently [13].  In particular, the
dilaton and the antisymmetric tensor fields, contained in what we
call SDTM [5], are expected to emerge {\it via} duality transformations
[13] from the original 70 scalar fields in the
$~N=8$~ theory.  Motivated by these
developments, it is also important to study more details of the $~N=1$~
SDSG, especially its {\it canonical} system, which contains an
irreducible {\it core} multiplet of the above truncated system.

        In the present paper, we will establish the formulation of
{\it canonical} superspace constraints, where the SD
especially for the
Riemann tensor is manifest, and the supertorsion vanishes.  In this
formulation, the significance of the SDTM we
have presented in a previous paper is elucidated.  We show that
the three multiplets of the SG, TM and SYM all
undergo SD conditions, which are mutually consistent with each other.
We stress that even in the string amplitude calculation [8], the
compatibility between the SDYM and SDG has {\it not} been obvious.  In
particular, the SD conditions get modified, when the SDYM is coupled to SDG.
Our present paper promotes the understanding about the mutual consistency
between SDSG and SDSYM in terms of superspace
formulation.\footnotew{It is well-known that amplitude calculation
always allows some room for field-redefinitions in the point field
theory limit.  We hope our canonical formulation provides helpful
information to understand the explicit consistency.}

        We also show that these self-dual gauge field backgrounds can
consistently couple to the GS superstring $~\s\-$model,
maintaining the fermionic $~\k\-$invariacne.  We will demonstrate
how the constraints onto the SD gauge fields are realized at the
lagrangian level of the GS $~\s\-$model.

        This paper is organized as follows.  In the next section, we
establish the $~D=(2,2),\,N=1$~ superspace constraints for the SG + TM
+ SYM.  In section 3, we inspect the consistency of our SD conditions
with supersymmetry, and get all the superfield equations under these SD
conditions.  In section 4, we construct a Green-Schwarz superstring
$~\s\-$model, which has $~\kappa\-$invariance in the presence of the
self-dual superspace backgrounds.  We also give the constraint
lagrangian, that gives the SD condition automatically.
Section 5 is devoted to concluding remarks.  In Appendix A, we give notational
remarks and some useful formulae for practical computations in our
$~D=(2,2),\,N=1$~ superspace.  Appendix B is for superfield equations
{\it before} imposing SD conditions, and in Appendix C we give strong
evidence that SDSG is finite to all-orders, based on reasonable
assumptions.  Appendix D is devoted to a set of
exact solutions for the coupled system of SDSG + SDSYM + SDTM on a self-dual
gravitational instanton background.

\bigskip\bigskip\bigskip

\centerline {\bf II.~~Canonical Constraints}
\medskip\medskip

        We first establish the {\it canonical} set of constraints, in
which the SD conditions will be manifest.  Our supermultiplets are the SG
$~(e\du\ula\ulm,\psi\du\ulm{\ul\a})$, the TM $~(B_{\ulm\uln},
\chi_{\ul\a}, \Phi)$~ and the SYM $~(A\du\ulm I, \l_{\ul\a}{}^I)$.
We use the notation compatible with our previous papers [2-4], i.e., the {\it
underlined} indices $~{\scst \ula,~\ulb,~\cdots}$~ stand for the
vectorial $~D=(2,2)$~ local Lorentz indices, and $~{\scst \ulm,~ \uln,~
\cdots}$~ are for the vectorial {\it curved} indices, while $~{\scst
\ul\a,~\ul\b,~ \cdots}$~ are local spinorial indices as abbreviations
for the respective set of indices $~{\scst (\a,\,\Dot\a),~(\b,\,\Dot\b),
{}~\cdots}$.  Relevantly, we use the {\it
tilded} superfields for the anti-chiral Weyl spinors.
The indices $~{\scst I,~J,~\cdots }$~ are for the adjoint
representations of the gauge group.

        Our constraints are to satisfy the Bianchi identities (BIds.)
$$\eqalign{&\nabla_{\[ A} T\du{ B C)} D - T\du{\[ A B|} E T\du {E| C)}
D - R\du{\[ A B C )} D \equiv 0 ~~, \cr
&\fracm 16 \nabla_{\[ A} G_{B C D)} - \fracm 14 T\du{\[ A B|} E G_{E|C
D)} + \fracm 1{4\sqrt3} F\du{\[ A B} I F\du{C D)} I \equiv 0 ~~, \cr
&\nabla_{\[A} F\du{ B C)} I - T\du{\[ A B| } D F\du{D| C)} I \equiv
0~~.  \cr }
\eqno(2.1) $$
As usual, we need the $~F^2\-$term in the
$~G\-$BId.~in order to have the consistent coupling of TM to SYM.
The special factor $~\sqrt3$~ in the $~F^2\-$term is to have the
canonical kinetic term for the gauge field in our invariant lagrangian
(2.4) below.  Accordingly, the superfield strength is defined by
$$G_{M N P} \equiv \half \partial_{\[ M} B_{N P) }
- \fracm1{2\sqrt3} \left[ F\du{\[ M N} I A\du {P )} I - 2 f^{I J K} A\du
M I A\du N J A\du P K \right] ~~.
\eqno(2.2) $$

        We now assign the engineering dimensions to our superfields as
$~\[e\du\ula\ulm\] = \[ B_{\ulm\uln} \] = \[ \Phi \] = [A\du\ulm I \] =
0,~~\[ \psi\du\ulm{\ul\a} \] = \[ \chi_{\ul\a} \] = \[ \l\du{\ul\a} I \]
= 1/2$.  Thus, e.g., $~\[T\du{\ula\ulb}{\ul\g} \] = 1/2,~
\[T\du{\ula\ulb}\ulc \] = 1,~ \[G_{\ul\a\ul\b\ulc}\] =
0,~\[G_{\ul\a\ulb\ulc}\] = 1/2,~\[F_{\ul\a\ulb}{}^I\]  = 1/2$~ and
$~\[F_{\ula\ulb}{}^I \] = 1$, {\it etc}.
Then our superspace constraints have the dimensions $~0\le d
\le 1$.  We do {\it not} impose any SD conditions, until we make it
sure that all of the SD conditions are consistent with these
constraints under supersymmetry.  Our constraints can be obtained
by performing super-Weyl
rescaling from the ones given in Ref.~[3], or even by direct
construction, putting some unknown coefficients, satisfying the
BIds.~(2.1).  Our result is
$$\eqalign{&T\du{\a\Dot\b} \ulc = i(\s^\ulc)_{\a\Dot\b} ~~, ~~~~
\nabla_{\ul\a} \Phi = - \fracm 1{\sqrt 2} \chi_{\ul\a} ~~, \cr
&G_{\a\Dot\b c} = -\fracm {{\sqrt3}i} 2 \, (\s_\ulc)_{\a\Dot\b}
\, e^{2\Phi/{\sqrt3}} ~~, ~~~~T\du {\ul\a \ulb} \ulc = 0~~,
{}~~~~ G_{\ul\a\ul\b\ul\g} = 0~~, \cr
&G_{\a \ulb \ulc} = \fracm 1{\sqrt2} (\s{\low{\ulb\ulc}} \chi)_\a
\,e^{2\Phi/{\sqrt3}} ~~, ~~~~
G_{\Dot\a \ulb\ulc} = \fracm 1{\sqrt2} (\s{\low{\ulb\ulc}}\Tilde\chi)
_{\Dot\a} \, e^{2\Phi/{\sqrt3}}~~, \cr
&\nabla_{\a}\Tilde\chi_{\Dot\b} = -\fracm i{\sqrt2}(\s^\ulc)_{\a\Dot\b}
\nabla_\ulc \Phi + \fracm i {6{\sqrt2}} e^{-2\Phi/{\sqrt3}}
(\s^{\ulc\uld\ule})_{\a\Dot\b}\, G_{\ulc\uld\ule}
-\fracm1{4{\sqrt6}} (\s^\ulc)_{\a\Dot\b}
(\l^I\s_\ulc\Tilde\l^I) ~~, \cr
&\Tilde\nabla_{\Dot\a} \chi{\low\b} = - \fracm i {\sqrt2}
(\s^\ulc)_{\a\Dot\b} \nabla_\ulc\Phi
- \fracm i {6{\sqrt2}} e^{-2\Phi/{\sqrt 3}} (\s^{\ulc\uld\ule})_{\b\Dot\a}
G_{\ulc\uld\ule}
- \fracm1{4\sqrt6}(\s^\ulc)_{\b\Dot\a} (\l^I \s_\ulc\Tilde\l^I) ~~, \cr
&\nabla_\a\chi_\b = - \fracm 1{4{\sqrt6}} C_{\a\b} \, (\chi\chi)
+\fracm 1{\sqrt6} C_{\a\b} (\Tilde\l^I \Tilde\l^I) ~~, \cr
&\Tilde\nabla_{\Dot\a} \Tilde\chi_{\Dot\b} = -
\fracm 1{4\sqrt6}C_{\Dot\a\Dot\b} (\Tilde\chi\Tilde\chi) + \fracm1{\sqrt6}
C_{\Dot\a\Dot\b} (\l^I \l^I) ~~, \cr
&T\du{\a\b}\g = - \fracm{\sqrt3}{2\sqrt2} \d\du{(\a} \g
\chi_{\b)} ~~, ~~~~
T\du{\Dot\a\Dot\b} {\Dot\g} = - \fracm{\sqrt3}{2{\sqrt2}}\du{ (\Dot\a} {\Dot\g}
\Tilde\chi_{\Dot\b)} ~~, \cr
&T\du{\a\Dot\b} {\Dot\g} = \fracm{\sqrt3}{2\sqrt2} \d\du{\Dot\b}{\Dot\g}
\chi_\a ~~, ~~~~ T\du{\Dot\a\b} \g = \fracm{\sqrt3} {2\sqrt2} \d\du\b\g
\Tilde\chi_{\Dot\a} ~~, \cr
&T\du{\a\b} {\Dot\g} = 0~~, ~~~~ T\du{\Dot\a\Dot\b} \g = 0~~, ~~~~
T\du{\ula \ulb} \ulc = 0~~, ~~~~ T\du{\a \ulb} {\Dot\g} = 0 ~~, ~~~~
T\du{\Dot\a\ulb} \g = 0~~, \cr
&T\du{\a\ulb} \g = + \fracm1{4\sqrt3} \,e^{-2\Phi/\sqrt3}\,
\e\du \ulb{\ulc\uld\ule} \,
\d\du\a\g G_{\ulc\uld\ule} - \fracm{3i}{16} \d\du\a\g (\chi\s_\ulb\Tilde\chi)
+\fracm{7i}{2\sqrt6} (\s_\ulb\Tilde\chi)_\a \chi^\g \cr
& ~~~~~ ~~~~~ ~~~~~
- \fracm i {12} \d\du\a\g (\l^I\s_\ulb\Tilde \l^I)
- \fracm i 6 (\s\du \ulb\ulc)\du \a\g (\l^I \s_\ulc \Tilde\l^I) ~~, \cr
&T\du{\Dot\a \ulb} {\Dot\g} = -\fracm1{4\sqrt3} \, e^{-2\Phi/\sqrt3}\, \e\du
\ulb{\ulc\uld\ule} \, \d\du{\Dot\a}{\Dot\g} G_{\ulc\uld\ule}
+ \fracm{3i} {16} \d\du{\Dot\a}{\Dot\g}
(\chi\s_\ulb\Tilde\chi) +\fracm{7i}{2\sqrt6} (\s_\ulb\chi)_{\Dot\a}
\Tilde\chi^{\Dot\g} \cr
& ~~~~~ ~~~~~ ~~~~~ + \fracm i {12} \d\du{\Dot\a}{\Dot\g}
(\l^I\s_\ulb\Tilde \l^I)
+ \fracm i 6 (\s\du \ulb \ulc)\du{\Dot\a}{\Dot\g}
(\l^I \s_\ulc \Tilde\l^I) ~~, \cr
&F\du{\a \ulb} I = \fracm i {\sqrt2}
e^{\Phi/\sqrt3} (\s_\ulb \Tilde\l^I )_\a ~~,
{}~~~~ F\du{\Dot\a \ulb} I = \fracm i{\sqrt2} e^{\Phi/\sqrt3}(\s_\ulb\l^I)
_{\Dot\a} ~~, ~~~~F_{\ul\a\ul\b} {}^I = 0 ~~, \cr
&\nabla_\a\l\du\b I = - \fracm1{2\sqrt2} (\s^{\ulc\uld})_{\a\b}\,
e^{-\Phi/\sqrt3} F\du{\ulc\uld} I + \fracm {\sqrt3} {16\sqrt2} (\s^{\ulc\uld})
_{\a\b} (\chi\s_{\ulc\uld} \l^I) +\fracm{\sqrt3} {4\sqrt2} C_{\a\b} (\chi\l^I +
\Tilde\chi\Tilde\l^I) ~~, \cr
&\Tilde\nabla_{\Dot\a}\Tilde\l\du{\Dot\b} I = - \fracm1{2\sqrt2} (\s^{\ulc
\uld})_{\Dot\a\Dot\b}\,
e^{-\Phi/\sqrt3} F\du{\ulc\uld} I + \fracm {\sqrt3} {16\sqrt2} (\s^{\ulc
\uld})_{\Dot\a\Dot\b} (\Tilde\chi\s_{\ulc\uld} \Tilde\l^I)
+\fracm{\sqrt3} {4\sqrt2} C_{\Dot\a\Dot\b} (\chi\l^I +
\Tilde\chi\Tilde\l^I) ~~, \cr
&\nabla_\a\Tilde\l^{\Dot\b\,I} = - \fracm1{ 2\sqrt6} \, \chi_\a
\Tilde\l^{\Dot\b\,I} ~~, ~~~~
\Tilde\nabla_{\Dot\a}\l^{\b\,I} = - \fracm1{ 2\sqrt6} \, \Tilde \chi_{\Dot\a}
\l^{\b\,I} ~~. \cr}
\eqno(2.3) $$
We are using the symbols like ~$\chi\chi \equiv
\chi^\g\chi_\g$, ~$\Tilde\chi\Tilde\chi \equiv \Tilde\chi^{\Dot\g}
\Tilde\chi_{\Dot\g}$, {\it etc}.~as explained in the Appendix A.  It is
to be emphasized that the
supertorsion $~T\du{\ula\ulb}\ulc $~ is {\it vanishing} in this set of
constraints.  This feature is important to guarantee that
of the SD condition on the Riemann tensor is a sufficient condition for the
Ricci-flatness of the space-time, as will be seen in the next section.

        The terminology ``canonical'' stems from the fact that the
invariant lagrangian for our system has the {\it canonical}
kinetic terms.  In fact, the invariant lagrangian is given
by\footnotew{In accord with the superspace notation [15], we use
$~e\equiv{\rm d e t}(e\du\ula\ulm)$, so that the lagrangian density
has the common factor  $~e^{-1}$~ as a scalar density.}
$$\eqalign{e \Lag = \,& -\fracm 14 R + \fracm 1{12}
e^{-4\Phi/\sqrt3} G_{\ula\ulb\ulc} ^2 + \half (\nabla_\ula\Phi)^2
-\fracm 14 e^{-2\Phi/\sqrt3} (F\du{\ula\ulb} I)^2 \cr
& - \fracm i2 \left(\psi_\ula \s^{\ula\ulb\ulc} \Tilde
T_{\ulb\ulc}\right) +
i \left(\chi\s^\ula \nabla_\ula \Tilde \chi\right) + i \left(\l^I \s^\ula
\nabla_\ula\Tilde\l^I\right) \cr
& + \fracm1{6\sqrt2} e^{-2\Phi/\sqrt3} \left( \psi_\ula \s^{\ulb\ulc\uld}
\s^\ula \chi + \Tilde \psi_\ula \s^{\ulb\ulc\uld} \s^\ula \Tilde \chi
\right) G_{\ulb\ulc\uld} \cr
&- \fracm 1{\sqrt2} \left( \psi_\ula \s^\ulb \s^\ula \chi
+ \Tilde \psi_\ula \s^\ulb\s^\ula \Tilde\chi \right) \nabla_\ulb \Phi
- \fracm{{\sqrt3}i}4 (\psi_\ula\s_\ulb\Tilde\psi_\ulc) G^{\ula\ulb\ulc}
\cr
&+ \fracm i {2\sqrt2} e^{-\Phi/\sqrt3} \left(\psi_\ula \s^{\ulb\ulc} \s^\ula
\Tilde\l^I + \Tilde \psi_\ula \s^{\ulb\ulc} \s^\ula \l^I \right) F\du{\ulb\ulc}
I \cr
&- \fracm 1{2\sqrt3} e^{-\Phi/\sqrt3} \left( \chi\s^{\ula\ulb}\l^I
+ \Tilde \chi\s^{\ula\ulb}\Tilde \l^I  \right) F\du{\ula\ulb} I \cr
&- \fracm i {12\sqrt3} e^{-2\Phi/\sqrt3} \left(\chi \s^{\ula\ulb\ulc}
\Tilde\chi + \l^I \s^{\ula\ulb\ulc} \Tilde\l^I \, \right)
e^{-2\Phi/\sqrt3} G_{\ula\ulb\ulc} ~~, \cr }
\eqno(2.4) $$
up to fermionic quartic terms.  The usefulness of this invariant
lagrangian manifests itself, when considering dimensional
reduction or compactifications into lower-dimensions [9].

        Due to the similarity of our $~D=(2,2),\,N=1$~ system to
$~D=(1,3),\,N=1$~ theories, we see that our constraints (2.3) share
the same magnitudes of coefficients with the $~D=(1,3)$~component
results in Ref.~[16], as far as the {\it non}-SYM sector is concerned.

\bigskip\bigskip\bigskip\bigskip

\centerline {\bf III.~~Self-Duality Conditions}
\medskip\medskip

        We are now ready to inspect the consistency of our SD conditions
on each multiplet with the constraints (2.3).  Our supersymmetric SD
conditions will be
$$\li{&T_{\ula\ulb}{}^\g= 0~~,
&(3.1) \cr
&\Tilde\chi_{\Dot\a} = 0 ~~,
&(3.2) \cr
&\l^{\a\, I} = 0~~,
&(3.3) \cr } $$
respectively on the SG, TM and SYM multiplets.\footnotew{We mention also that
(3.1) is stronger than the condition $~W_{\a\b\g} = 0$~ for the {\it
purely} SDSG case in Refs.~[3,5]
for a reason mentioned in Appendix C.  After these SD conditions are
imposed, these multiplets are respectively called SDSG, SDTM and SDSYM.}~~These
conditions are sufficient for the bosonic SD conditions
$$\li{&R\du{\ula\ulb}{\ulc\uld} = \half \e \du{\ula\ulb}{\ule\ulf}
R\du{\ule\ulf}{\ulc\uld} ~~,
&(3.4) \cr
&G_{\ula\ulb\ulc} = e^{2\Phi/{\sqrt3}}\, \e\du{\ula\ulb\ulc}\uld
\,\nabla_{\uld} \Phi ~~,
&(3.5) \cr
&F\du{\ula\ulb} I = \half \e\du{\ula\ulb}{\ulc\uld} F\du{\ulc\uld} I ~~.
&(3.6) \cr } $$
Eq.~(3.5) is an analog of the usual SD
condition for the SDYM, connecting the three form $~G_{\ula\ulb\ulc}$~ with the
one-form $~\nabla_{\ula}\Phi$.  For this reason we use the word ``SDTM''.

        Our next task is to confirm the consistency of the SD
conditions (3.1) - (3.6) with the constraints (2.3).  First of all, we
see that the SYM sector is simple, because all the fermionic bilinear
terms disappear, leaving only the $~F\-$term in $~\Tilde\nabla_{\Dot\a}
\Tilde\l^{\Dot\b\, I}$~ exactly as the {\it globally} supersymmetric case
in Ref.~[2], while $~\Tilde\nabla_{\Dot\a}\l^{\b\, I}$,
$~\nabla_{\a}\l^{\b\, I}$ are now vanishing.  This assures the
consistency of the SD condition on the SYM.
The consistency of (3.2) on the purely TM is also easily seen,
especially the ``SD condition'' on the TM is compatible with the
equation $~\nabla_\a \Tilde \chi_{\Dot\b}=0$.  Actually using $~\l\du \a I
= 0$~ (3.3), we have
$$\nabla_\a \Tilde\chi_{\Dot\b} = - \fracm i2 (\s^\ulc )_{\a\Dot\b}
\left( \nabla_\ulc \Phi - \fracm 16 \e\du\ulc{\uld\ule\ulf}
e^{-2\Phi/\sqrt3} G_{\uld\ule\ulf} \right) ~~,
\eqno(3.7) $$
which vanishes under (3.5).  The equation $~\Tilde\nabla_{\Dot\a}
\Tilde\chi_{\Dot\b} = 0$~ is trivially satisfied under our SD conditions
(3.1) - (3.3).  This is essentially the
same as the global case demonstrated in Ref.~[5].
The consistency on the
purely SG sector has been already explained in Ref.~[3], so that the
question is the {\it mixing} between the SG and TM or SYM, which had not been
accomplished in our previous papers [3,5].   In our present system, this
is easy to realize, because all the effect by the TM is consistently deleted,
once the SD conditions (3.1) - (3.3) are imposed.

        We also mention that the above SD conditions can be
completely replaced by the {\it anti-self-duality} conditions, by
switching the chiralities of all the Weyl spinors in (3.1) - (3.3).
However, it is important that these conditions are to be {\it
supersymmetrically} consistent with each
other.  For example, we can {\it not} couple the SDSG to {\it
anti-self-dual} YM multiplet, {\it etc}.  This feature is
did {\it not} exist in a {\it non-supersymmetric} self-dual gauge
field system.

        The superfield equations are as usual obtained by utilizing the
BIds.~at $~d\ge 3/2$.  Once we have understood the consistency of our
SD conditions (3.1) - (3.6), we can impose them on the relevant
superfields, when getting the superfield equations.
Our superfield equations for our {\it self-dual} multiplets: SDSG + SDTM
+ SDSYM thus obtained are
$$\li{& i (\s^\ulb)_{\a\Dot\b}\Tilde T\du{\ula \ulb} {\Dot\b} = 0 ~~,
&(3.8) \cr
&i (\s^\ula)\ud\b{\Dot\a} \nabla_\ula \chi{\low\b}
+ \fracm i {2\sqrt3} (\s^\ula\chi)
_{\Dot\a} \nabla_\ula\Phi - \fracm1{2\sqrt3} (\s^{\ula\ulb}\Tilde\l^I)_{\Dot\a}
e^{- \Phi/\sqrt3} F\du{\ula\ulb} I = 0~~,
&(3.9) \cr
& i(\s^\ula)\du\a{\Dot\b} \nabla_\ula \Tilde\l _{\Dot\b}{}^I
- \fracm i{2\sqrt3} (\s^\ula \Tilde\l^I)_\a \nabla_\ula \Phi = 0 ~~,
&(3.10) \cr
& R\du{\ula\ulb}{\ulc\uld} = \half \e \du{\ula\ulb}{\ule\ulf}
R\du{\ule\ulf}{\ulc\uld} ~~,~~~~R_{\ula\ulb} = 0~~,
&(3.11) \cr
& \Bo \Phi + \fracm2{\sqrt3} (\nabla_\ula\Phi)^2 - \fracm1{2\sqrt3}
e^{-2\Phi/\sqrt3} (F_{\ula\ulb}{}^I)^2 = 0~~, ~~~~
G_{\ula\ulb\ulc} = e^{2\Phi/{\sqrt3}}\, \e\du{\ula\ulb\ulc}\uld
\,\nabla_{\uld} \Phi ~~,
&(3.12) \cr
& F\du{\ula\ulb} I = \half \e\du{\ula\ulb}{\ulc\uld} F\du{\ulc\uld} I
{}~~, ~~~~ \nabla_\ula F^{\ula\ulb\, I} = 0 ~~.
&(3.13) \cr} $$

Notice that the superfield equations (3.8) and (3.11) are {\it not}
affected either by the TM or the SYM.  This is closely related to the
fact that the Ricci-flatness (3.11) should {\it not} be disturbed by the
energy-momentum tensor of ``matter'' fields.  In fact, we can check this by
looking into the $~d=2$~ BId.~$\nabla_{\[\ula} T\du{\ulb\ulc\]} \uld
- T\du{\[\ula\ulb|} \ule
T\du{\ule | \ulc\]} \uld - T\du{\[ \ula\ulb|}{\ul\e} T\du {\ul\e| \ulc\]
} \uld - R\du {\[\ula\ulb\ulc\]} \uld \equiv 0$
which implies
$$R_{\[\ula\ulb\ulc\]\uld} = 0~~,
\eqno(3.14) $$
because of our constraints
$~T\du{\ula\ulb}\ulc = 0$, ~$T\du{\ul\a\ulb}\ulc = 0$.  This in turn
means the Ricci-flatness, since
$$R_{\ula\ulb} = R\du{\ula\ulc\ulb}\ulc = \half\e\du{\ula\ulc}{\uld\ule}
R\du{\uld\ule\ulb}\ulc = - \fracm 1{12} \e\du\ula{\ulc\uld\ule} R_{\[
\ulc\uld\ule\]\ulb} = 0~~.
\eqno(3.15) $$
Therefore the energy-momentum tensor in the gravitational field equation
is to {\it vanish!}
The absence of the source term for
the YM in (3.13) is of the same sort.  The $~\chi,~\Phi$~ and
$~\l\-$field equations feel some effects of other multiplets, as seen in
(3.9), (3.10) and (3.12).  Recall also that $~G_{\ula\ulb\ulc}$~ in
(3.12) contains the CS-piece as in (2.2).

        We finally have to check the consistency of these field
equations under our SD conditions with the $~d=5/2$~ BIds.  This is
actually not difficult, and we give here a typical example.  We can take
the spinorial derivative of eq.~(3.4), which is to vanish:
$$\eqalign{\nabla_\a & \left[\, R\du{\ulb\ulc}{\uld\ule} - \half
\e\du{\ulb\ulc}{\ulf\ulg} R\du{\ulf\ulg}{\uld\ule} \,\right] \cr
& = \fracm i2 e^{-{\sqrt3}\Phi /2} \bigg[ \left( \s^{\[ \uld}
\nabla_{\[ \ulb} {\breve T}\ud{\ule \]}{\ulc \]} - \s_{\[ \ulb}
\nabla_{\ulc\]} {\breve T}^{\uld\ule} \right)_\a   \cr
&~~~ ~~~~~ ~~~~~ ~~~~~ - \half \e\du{\ulb\ulc}{\ulf\ulg} \left( \s^{\[ \uld}
\nabla_{\[ \ulf} {\breve T} \ud{\ule \]} {\ulg\]} + \s_{\[ \ulf}
\nabla_{\ulg\]} {\breve T}^{\uld\ule} \right)_\a \bigg] ~~, \cr }
\eqno(3.16) $$
where $~{\breve T}\du{\ula\ulb}{\Dot\g}\equiv e^{{\sqrt3}\Phi/2}
\Tilde T \du{\ula\ulb}{\Dot\g}$.  Now the $~d=5/2$~ BId.
$$\nabla_{\[ \ula} \Tilde T \du{\ulb\ulc\]} {\Dot\d} - \Tilde
T\du{\[\ula\ulb|} {\Dot\e} T\du {\Dot\e| \ulc\]} {\Dot\d}
= e^{-{\sqrt3}\Phi/2} \nabla_{\[\ula} {\breve T} \du {\ulb\ulc\]}
{\Dot\d} \equiv 0~~,
\eqno(3.17) $$
helps us to rewrite (3.16) as
$$\nabla_\a \left[\, R\du{\ulb\ulc}{\uld\ule} - \half
\e\du{\ulb\ulc}{\ulf\ulg} R\du{\ulf\ulg}{\uld\ule} \, \right]
= - \fracm i2 (\s_{\ulb\ulc})\du \a{\Dot\g} \nabla^{\[ \uld} (\s_\ulg
{\breve T}^{\ule \]\ulg})_{\Dot\g} = 0 ~~,
\eqno(3.18) $$
since $~(\s^\ulb \breve T_{\ula\ulb})_\a = e^{{\sqrt3}\Phi/2}
(\s^\ulb\Tilde T_{\ula\ulb})_\a = 0$~ by (3.8).  This final check
guarantees the total consistency of our SD conditions with
supersymmetry, satisfying {\it all} the BIds.

        We also mention that the superfield equations (3.8) -
(3.13) can be re-obtained by taking variations of our invariant
lagrangian (2.4), and imposing afterwards the SD conditions (3.1) -
(3.6) on these field equations.  The usefulness of the invariant
lagrangian (2.4) manifests itself, when we want the general
field equations {\it before} imposing the SD conditions.  As is already
well known as the general feature of self-dual systems, the invariant
lagrangian (2.4) vanishes up to a total divergence, if the SD conditions
are directly imposed on itself.

\bigskip\bigskip\bigskip

\centerline {\bf IV.~~Green-Schwarz String $~\s\-$Model}
\medskip\medskip

It has been well known in the string amplitude calculation [8]
that the {\it bosonic} SDG + self-dual YM gauge fields
can be the consistent background for the $~N=2$~ superstring.  From this
viewpoint, it is natural expectation that these {\it supersymmetric}
target space-time backgrounds can be the consistent backgrounds for a
GS superstring.  In fact, we have presented such $~\s\-$model
in our previous paper [3] for the BFFC.
In the present paper, we construct the GS string coupled to these {\it
canonical} backgrounds, and show the invariance under the {\it fermionic}
$~\k\-$symmetry, which is required for the GS formulation for its consistency.

As has been mentioned in the Introduction, our GS formulation for
$~N=1$~ target space-time supersymmetry corresponds to some truncation
of the more consistent $~N=2$~ GS superstring developed by Siegel [13].
In this sense, our results below will still be of importance in the
context of GS $~\s\-$model formulation.

Our total action is [3]
$$I_{\rm GS} = I_{\rm SG} + I_{\rm YM} + I_{\rm C} ~~,
\eqno(4.1) $$
where\footnotew{There is to be an overall factor $~1/(2\pi)$~ in front
of these integrals, to accord with instanton solutions given in
Ref.~[10], which we omit universally {\it except for} Appendix D.}
$$\li{&I_{\rm SG} = \int d^2 \s \,\left[\,\half V^{-1} g^{i j}
\eta_{\ula\ulb} \, e^{2\Phi/\sqrt3} \, \Pi\du i\ula \Pi\du j\ulb
- \fracm1{\sqrt3} \e^{i j} \Pi\du i A \Pi \du j B B_{B A} \,\right]
{}~~,
&(4.2) \cr
&I_{\rm YM} = \int d^2 \s \, \fracm i2 V^{-1} \Bar\psi \du
- r \g^i \left[\, \partial_i
\psi\du - r + \Pi \du i A A\du A I (T_I \psi_-)^r \right] ~~,
&(4.3) \cr } $$
and $~I_{\rm C}$~ is a constraint action to be described shortly.
As usual, $~Z^M$~ are the target superspace coordinates, and $~\Pi\du i
A\equiv (\partial_i Z^M) E\du MA$.  We use the indices $~{\scst
i,~j,~\cdots}$~ for the world-sheet curved coordinates.  The spinors
$~\psi\du - r$~ are unidexterous, and are in an appropriate representation
of a YM gauge group, whose generators are $~(T_I)_r{}^s$.  The $~V$~ is the
determinant of the zweibein on the world-sheet.
The appearance of the special factor $~1/{\sqrt3}$~ is due to our
peculiar constraint in $~G_{\a\Dot\b \ulc}$~ in (2.3).

We can confirm that the action (4.2) + (4.3) is invariant under the
{\it fermionic} $~\k\-$transformation
$$\eqalign{&\d_\k E^{\ul\a} = i (\s_\ula)^{\a\Dot\b} \Pi\du\plpl \ula \,
\Tilde\k_{\mimi\Dot\b} + i (\s_\ula)^{\b\Dot\a} \Pi\du\plpl \ula
\k_{\mimi\b} ~~, \cr
&\d_\k V\du \mimi i = 4 \left[ (\k_\mimi\Pi_\mimi) + (\Tilde\k_\mimi
\Tilde\Pi_\mimi) \right] V \du \plpl i
- {\sqrt{\fracm 23}}i \left[ (\k_\mimi \s_\ula
\Tilde\chi) + (\Tilde\k_\mimi \s_\ula \chi) \right] \Pi\du \mimi
\ula V\du \plpl i \cr
& ~~~~~ ~~~~~  - \fracm i{2{\sqrt2}} e^{\Phi/\sqrt3} V \du \plpl i
(\Bar\psi_- T^I \psi_- ) (\k_\mimi\l^I + \Tilde\k_\mimi\Tilde\l^I) ~~,
\cr
&\d_\k V\du \plpl i = 0 ~~, ~~~~\d_\k E^\ula = 0 ~~, \cr
&\d_\k \psi\du - r = - (\d E^{\ul\a}) A\du{\ul\a} I (T_I\psi_-) ^r ~~, \cr }
\eqno(4.4) $$
required for classical consistency in GS string $~\s\-$model.
We are using $~\d E^A \equiv (\d Z^M) E\du M A$, while $~{}_\plpl$~ and
$~{}_\mimi$~ are for the light-cone
projections: $~(1/2) (\eta^{i j} \pm V^{-1} \e^{i j})$.  The $~V \du\plpl
i$~ and $~V \du\mimi i$~ are the zweibeins.  The invariance check above
is easily performed in a similar way to
Refs.~[3,12].  The difference of our {\it canonical} case from the BFFC
in Ref.~[3] is recognized by the appearance of the new term with
$~\chi$~ and the exponential factor $~\exp(2\Phi/{\sqrt3})$.

        We now address ourselves to the question of the {\it constraint}
lagrangians producing our SD conditions (3.1) - (3.3).
We have already given some
clue in Ref.~[3], concerning the SDYM multiplet.
In this paper we give a more complete result, which is made possible
due to the manifest consistency of our SD conditions (3.1) - (3.6).
Our constraint lagrangians are
$$\li{&I_{\rm C} \equiv I_{\rm C, \, SG} + I_{\rm C, \, TM} + I_{\rm C,
\, YM} ~~, \cr
&I_{\rm C, \, SG} \equiv \int d^2 \s \, V^{-1} \left[\,
\Pi\du\plpl\ula \Pi\du \mimi\ulb
(\nabla_\plpl\Pi_{\mimi\g}) T_{\ula\ulb}{}^\g \, \right] ~~,
&(4.5) \cr
& I_{\rm C,\, TM} \equiv \int d^2 \s \,  V^{-1} \left[\,
i \Pi\du \plpl \a \Pi\du
\mimi \ula (\s_\ula)\du \a{\Dot\b} \Tilde\chi_{\Dot\b} \,\right] ~~,
&(4.6) \cr
&I_{\rm C,\, YM} \equiv \int d^2 \s \, V^{-1} \left[ \,\Pi\du\plpl\a \l_\a{}^I
\,\right] i (\Bar\psi_-T_I\psi_-)~~.
&(4.7) \cr } $$
In particular, the special derivative factor in (4.5) is chosen such
that the integrand is {\it not} a total divergence.
and all the $~{}_\plpl$~ and $~{}_\mimi$~ appear only
in pairs in order to guarantee the $~D=2$~ Lorentz invariance.

We now discuss the $~\k\-$invariance of $~I_C$.
We easily see that all the terms in the
$~\k\-$variation of $~I_{\rm C,\, SG}$ $~I_{\rm C,\, TM}$~ and
$~I_{\rm C,\, YM}$~ are {\it proportional} to the SD conditions
(3.1) - (3.3).  In fact, using the Lorentz connection superfield
$~\phi\du{A B} C$~ in $~D=(2,2)$, we get
$$\eqalign{&\d_\k T\du{\ula\ulb}\g = (\d_\k E^{\ul\d}) \left[\, \nabla_{\ul\d}
T\du{\ula\ulb}\g - \phi\du{\ul\d \[ \ula |}\ulc T\du{\ulc| \ulb\]}\g
- \phi\du{\ul\d}{\g\e} T_{\ula\ulb \e} \,\right] ~~, \cr
&\d_\k \Tilde\chi_{\Dot\a} = (\d_\k E^{\ul\g}) \, \left[\,
\nabla_{\ul\g} \Tilde\chi_{\Dot\a} - \phi\du{\ul\g \Dot\a} {\Dot\b}
\Tilde\chi_{\Dot\b} \, \right] ~~, \cr
&\d_\k \,\l_\a{}^I  = (\d_\k E^{\ul\g} )
\left[ \nabla_{\ul\g} \l\du\a I - \phi\du{\ul\g\a}\d \l\du\d I
- f\du{J K}I A\du {\ul\g} J \l\du\a K \right] ~~. \cr}
\eqno(4.8) $$
so that the $~\k\-$transforms of $~\Tilde\chi\du{\Dot\a} I,~\l\du\a I$~ and
$~T\du{\ula\ulb}\g$~ are all {\it proportional to themselves}.
Therefore the total action $~I_{\rm SG} +
I_{\rm YM} + I_{\rm C}$~ is
$~\k\-$invariant, {\it only when} the background superfields satisfy the SD
conditions (3.1) - (3.3).  Thus our ~$\k\-$invariance restricts the
backgrounds to be self-dual supersymmetric gauge fields.  We can consider also
the constraint lagrangian for the {\it off-shell} SG, as given in Appendix C.

\vfill\eject

\centerline {\bf V.~Concluding Remarks}
\medskip\medskip

        We have constructed the {\it canonical} set of
constraints, which reveals the manifest consistency of the SD
conditions with supersymmetry.  In the present paper we have also
shown that such backgrounds
can be coupled to the GS string $~\s\-$model, maintaining
the $~\k\-$invariance.
We have shown that the SD conditions are automatically implied by
the constraint term $~I_{\rm C}$~ in our total GS action.

        In our previous paper [3], the manifestation of the consistency
of SD condition with supersymmetry was not well elucidated, due to the
non-vanishing supertorsion $~T\du{\ula\ulb} \ulc$.  In particular, the
consistency of the Ricci-flatness with the gravitational equation needed
some additional field redefinitions.  In the present paper, we have
circumvented this problem
by using the {\it canonical} constraints, where the Ricci-flatness is
manifestly equivalent to the self-dual Riemann tensor.  Our system is
useful in practical applications of the SDSG and SDYM system, especially
when we need to look into the interactions between them.
We believe that our results have verified the consistent
couplings among these supersymmetric self-dual gauge multiplets in a
manifest way, for the first time in
the literature.

        Even though we have not performed the $~\b\-$function
calculation in the present paper, its validity has been already guaranteed.
This is
because we know that the super-Weyl rescaling (superfield redefinitions) [17]
does {\it not} change the physical significance of the system.

        We have not introduced any {\it auxiliary fields} so far, which
are useful in the usual {\it off-shell} couplings of matter multiplets.
In the case
of self-dual supersymmetry, such auxiliary fields do not play
any significant role, because the SD is essentially the {\it on-shell}
concept, as has been also mentioned in our previous paper [3].  However,
in the Appendix B we have shown the most important application of the auxiliary
fields in the context of the proof that all the string corrections to the GS
string $~\b\-$functions for the SDYM backgrounds are to vanish.

        We have dealt only with self-dual multiplets in this
paper.  The remarkable point is that in $~D=(2,2)$~ the ``chirality'' of
fermions in each multiplet {\it automatically} implies the SD (or anti-SD)
conditions on the multiplet consistently with supersymmetry.
(We have seen this is true even with the TM.)  This feature seems common
to all the dimensions of $~D=(2+2n,2+2n)\-$type with an arbitrary
integer $~n$, where the Clifford algebra structure are similar [5,18].
Thus we expect similar situation in $~D=(4,4)$~ for SDTM with the
field strength of rank 4.

        In this paper, we have presented rather technical aspects of
self-dual gauge theories.  However, we expect that the results and
technique developed in this paper
will be useful for practical application of SDSG + SDTM + SDSYM, which
can be also the consistent backgrounds for the GS string.
In fact, we have utilized our canonical set of field equations for
obtaining exact solutions in Appendix D and in Ref.~[10].
Our results also provide an important technical and practical working-ground
for studying various {\it soluble} systems in lower-dimensions, as done
in our recent paper [9].

\bigskip\bigskip\bigskip

The author is grateful to S.J.~Gates, Jr.~for reading the manuscript and
giving important suggestions.  Acknowledgement is also due to
D.~Depireux, T.~H{\" u}bsch, T.~Jacobson, K.~Pirk and W.~Siegel
for fruitful discussions.

\vfill\eject

\centerline {\bf Appendix A: Notations and Useful Identities}
\medskip\medskip

In this appendix we give convenient lists of useful identities needed in
superspace computations.  This list is complimentary to the notational
appendix in our previous papers [2-5], and we avoid the repetition of the same
formulae in the latter.

We first give the Fierz rearrangement formulae, which are important in
superspace:
$$\eqalign{&(\s^\ula)_{\a\Dot\b} \, (\s_\ula)_{\g\Dot\d} = - 2 C_{\a\g}\,
C_{\Dot\b\Dot\d} ~~, ~~~~
(\s^\ula)^{\a\Dot\b} \,(\s_\ula)^{\g\Dot\d} = - 2C^{\a\g} \,
C^{\Dot\b\Dot\d} ~~, \cr
&(\s^\ula)_{(\a| \Dot\b} \, (\s_\ula)_{|\g) \Dot\d} = 0 ~~,
{}~~~~ (\s^{\ula\ulb})_{\a\b} (\s_{\ula\ulb})_{\g\d} = 4 C_{(\a|\g}
C_{|\b)\d} ~~.  \cr}
\eqno(A.1) $$
We can conveniently define
$$\li{&(\s^{\ula\ulb})\du\a\b \equiv \half (\s^{\[ \ula |} )\du\a{\Dot\g}
(\s^{| \ulb\]})\du{\Dot\g} \b ~~, ~~~~
(\s^{\ula\ulb})\du{\Dot\a}{\Dot\b} \equiv \half (\s^{\[ \ula |} )
\du{\Dot\a}\g (\s^{| \ulb \]} )\du\g{\Dot\b} ~~, \cr
&(\s^{\ula\ulb\ulc})\du\a{\Dot\b} \equiv \fracm 16 (\s^{\[ \ula |})
\du\a{\Dot\g} (\s^{|\ulb|})\du{\Dot\g}\d (\s^{| \ulc\]} )\du\d{\Dot\b}
{}~~, ~~~~(\s^{\ula\ulb\ulc})\du{\Dot\a}\b \equiv\fracm 16
(\s^{\[ \ula| } )\du{\Dot\a}\g
(\s^{|\ulb|})\du\g{\Dot\d} (\s^{|\ulc\]} )\du{\Dot\d}\b ~~,
&(A.2) \cr
&(\s^{\ula\ulb\ulc\uld})\du\a{\Dot\b} \equiv \fracm 1{24} (\s^{\[ \ula|}
)\du\a{\Dot\g} (\s^{|\ulb|})\du{\Dot\g} \d (\s^{|\ulc|})\du\d {\Dot\e}
(\s^{ |\uld\] }) \du {\Dot\e}\b ~~, ~~~~
(\s^{\ula\ulb\ulc\uld})\du{\Dot\a}{\Dot\b} \equiv \fracm 1{24}
(\s^{\[ \ula| } )\du{\Dot\a}\g (\s^{|\ulb|})\du\g{\Dot\d}
(\s^{|\ulc|})\du{\Dot\d} \e (\s^{ |\uld\] }) \du\e{\Dot\b} ~~. \cr}$$
As usual, the raising/lowering of the spinorial indices are performed by
$~C_{\a\b},~C^{\a\b},~C_{\Dot\a\Dot\b}$~ and $~C^{\Dot\a\Dot\b}$~, e.g.~$
(\s^\ulc)\du \a{\Dot\b} \equiv C^{\Dot\b\Dot\g} (\s^\ulc)_{\a\Dot\g}$,
{\it etc}.  In our notation in this paper, we consistently avoid
the expressions such
as the r.h.s.~of $~\{ \nabla_\a,\,\Tilde\nabla_{\Dot\b} \} =
i\nabla_{\a\Dot\b}$~ found in Ref.~[15].  Instead, we always put the
$~\s\-$matrices explicitly.
Additionally we universally {\it omit} the ``tilde''-symbols [2-5] on the
$~(\s^\ulc)_{\Dot\a\b}\-$matrices in this paper, because we can always
distinguish them from $~(\s^\ulc)_{\a\Dot\b}$~ by the ``{\it
dottedness}'' of their indices $~{\scst \Dot\a\b}$~ or by the
``{\it tildedness}'' of fermions multiplied by them.
Once this has been understood, we can take the advantage of this
compact notation, omitting lots of spinorial indices.  For example,
$$\eqalign{& (\s^\ula\s^{\ulb\ulc} \Tilde T_{\ulb\ulc} )_\a \equiv
(\s^\ula)\du\a{\Dot\b} (\s^{\ulb\ulc}) \du{\Dot\b}{\Dot\g} \Tilde
T_{\ulb\ulc\Dot\g} ~~, \cr
&\chi\chi \equiv \chi^\a\chi_\a~~, ~~~~ \chi\s^\ula\Tilde\l^I \equiv
\chi^\a (\s^\ula)\du\a{\Dot\b} \Tilde\l\du{\Dot\b} I~~, \cr
&\chi_\a\chi\low\b = -\fracm12 C_{\a\b} \, \chi\chi~~, ~~~~
\Tilde\chi_{\Dot\a}\Tilde\chi_{\Dot\b}
= - \half C_{\Dot\a\Dot\b} \, \Tilde\chi\Tilde\chi ~~. \cr}
\eqno(A.3) $$
Especially the supertorsion component $~\Tilde T\du{\ula\ulb}{\Dot\g}$~
is regarded as the {\it supercovariant} gravitino field
strength.  The most frequent formulae are the $~\s\-$mtarices with the
$~\e\-$tensors:
$$\eqalign{&(\s^{\ula\ulb})_{\a\b} = - \half \e^{\ula\ulb\ulc\uld}
(\s_{\ulc\uld})_{\a\b} ~~,  ~~~~
(\s^{\ula\ulb})_{\Dot\a\Dot\b} = + \half \e^{\ula\ulb\ulc\uld}
(\s_{\ulc\uld})_{\Dot\a\Dot\b} ~~, \cr
&(\s^{\ula\ulb\ulc})_{\a\Dot\b} = + \e^{\ula\ulb\ulc\uld}
(\s_\uld)_{\a\Dot\b} ~~, ~~~~
(\s^{\ula\ulb\ulc})_{\Dot\a\b} = - \e^{\ula\ulb\ulc\uld}
(\s_\uld)_{\Dot\a\b} ~~, \cr
&(\s^{\ula\ulb\ulc\uld}) _{\a\b} = + \e^{\ula\ulb\ulc\uld} \, C_{\a\b}
{}~~, ~~~~ (\s^{\ula\ulb\ulc\uld}) _{\Dot\a\Dot\b} = - \e^{\ula\ulb\ulc\uld} \,
C_{\Dot\a\Dot\b} ~~. \cr }
\eqno(A.4) $$
Relevantly, we have the useful relations
$$\e_{\ula\ulb\ulc\uld} \, \e^{\uld\ule\ulf\ulg} = - \d\du{\[ \ula} \ule
\, \d\du\ulb\ulf \d \du{\ulc\]} \ulg ~~, ~~~~
\e_{\ula\ulb\ulc\uld} \, \e^{\ulc\uld\ule\ulf} = 2 \d\du{\[ \ula} \ule
\, \d \du{\ulb\]} \ulf ~~, ~~~~
\e_{\ula\ulb\ulc\uld} \, \e^{\ule\ulb \ulc\uld} = 6 \d\du \ula\ule ~~.
\eqno(A.5) $$
It is also useful to have the relations
$$\eqalign{&(\s^{\ula\ulb})_{\a\b} \,(\s_{\ula\ulb} )
_{\Dot\g\Dot\d} \equiv 0 ~~, \cr
&(\s^{\[\ule |})_{\a\Dot\b}\,(\s^{|\ulf\]} )_{\g\Dot\d} = - C_{\a\g}
(\s^{\ule\ulf})_{\Dot\b\Dot\d} - C_{\Dot\b\Dot\d} (\s^{\ule\ulf})_{\a\g} \cr
&(\s^{\ula\[ \ulb|})_{\a\b} \, (\s^{|\ulc\] \ula})_{\Dot\g\Dot\d} \equiv
0 ~~,  \cr
&(\s^{\[ \ule}\s_{\ulg\ulh}\s^{\ulf\]} )_{\a\b} = 2 \left[\,
\d\du\ulg{\[\ule} \d\du\ulh{\ulf\]} + \e\du{\ulg\ulh}{\ule\ulf}
\,\right] C_{\a\b} ~~, \cr
&(\s^{\[ \ule} \s_{\ulg\ulh}\s^{\ulf\]} )_{\Dot\a\Dot\b} = 2
\left[ \, \d\du\ulg{\[\ule} \d\du\ulh{\ulf\]} - \e\du{\ulg\ulh}{\ule\ulf}
\,\right] C_{\Dot\a\Dot\b} ~~,  \cr
&(\s^\ula\s_{\ulb\ulc} \s_\ula)_{\a\b} = 0 ~~, ~~~~
(\s^\ula\s_{\ulb\ulc} \s_\ula)_{\Dot\a\Dot\b} = 0 ~~, \cr
&(\s^{\ula\ulb}\s_\ulc\s_{\ula\ulb})_{\a\b} = 0 ~~, ~~~~
(\s^{\ula\ulb}\s_\ulc\s_{\ula\ulb})_{\Dot\a\Dot\b} = 0 ~~. \cr }
\eqno(A.6)  $$
Any arbitrary {\it symmetric} matrices $~A_{\a\b}$~ and
$~S_{\Dot\a\Dot\b}$~ satisfy
$$A_{\a\b} \equiv \fracm18 (\s_{\ulc\uld})_{\a\b} (\s^{\ulc\uld})^{\g\d}
A_{\g\d}~~, ~~~~
S_{\Dot\a\Dot\b} \equiv \fracm18 (\s_{\ulc\uld})_{\Dot\a\Dot\b}
(\s^{\ulc\uld})^{\Dot\g\Dot\d} S_{\Dot\g\Dot\d} ~~,
\eqno(A.7) $$
which are equivalent to the usual identification like $~A_{\a\b} \approx
A_{\ula\ulb}$~ in Ref.~[15].  An interesting identity is for a
{\it self-dual} tensor $~S_{\ula\ulb}$~ and a {\it anti-self-dual} tensor
{}~$A_{\ula\ulb}$:
$$S^{\ulc\[ \ula} A\ud{\ulb\]} \ulc \equiv 0 ~~~~~
\bigg(\hbox{for~~} S_{\ula\ulb}
= \half \e\du{\ula\ulb}{\ulc\uld} S_{\ulc\uld} ~~, ~~~~
A_{\ula\ulb} = - \half \e\du{\ula\ulb}{\ulc\uld} A_{\ulc\uld}~\bigg) ~~.
\eqno(A.8) $$

\vfill\eject

\centerline{\bf Appendix B: Superfield Equations before SD Conditions}
\medskip\medskip

In this appendix, we give the superfield equations of SG + TM + YM, up
to trilinear fermionic terms {\it before} imposing
any SD conditions.  To get the
superfield equations in superspace is rather messy, but fortunately we
can utilize our invariant lagrangian (2.4).  To this end, we also use
the usual technique combining the $~\Phi\-$field equations with the
gravitational one, based on the {\it global scale covariance} of our lagrangian
(2.4):
$$\eqalign{&\Lag \rightarrow e^{-2c /\sqrt3} \, \Lag ~~, ~~~~
\Phi \rightarrow \Phi + c~~, ~~~~
e\du\ulm\ula \rightarrow
e^{-c/\sqrt3} e\du\ulm\ula ~~, \cr
&\psi\du\ulm{\ul\a} \rightarrow e^{-c/(2\sqrt3)}\, \psi\du\ulm{\ul\a} ~~,
{}~~~~ (\chi_{\ul\a} , \l\du {\ul\a} I) \rightarrow e^{c/(2\sqrt3)}\,
(\chi_{\ul\a} , \l\du {\ul\a} I) ~~, \cr }
\eqno(B.1) $$
with an arbitrary constant parameter $~c$.
Skipping all the details, which is essentially the same as eqs.~(4.11) -
(4.16) in Ref.~[19], we see that the gravitational equations is equivalent to
$$\fracmm{\d \Lag } {\d e\du\ulm\ula} = e^{-1}
\fracmm {\d L} {\d e\du\ulm\ula} - \fracm{\sqrt3} 2 e\du\ula\ulm
\, \partial_\uln \left[ \fracmm{\partial \Lag}
{\partial(\partial_\uln\Phi) }\right] = 0 ~~,
\eqno(B.2) $$
where $~L \equiv e\Lag$.
This implies that we do {\it not} have to take the variation of the
$~e \equiv {\rm d e t} (e\du\ula\ulm)$~ in the lagrangian, which is
conveniently replaced by the $~\Phi\-$field equation.  Similarly for the
YM-field equation, we recall the fact that the equation
$$\fracmm{\d \Lag }{\d A_\ulm} + \fracm2{\sqrt3} A_\uln \fracmm
{\d\Lag} {\d B_{\ulm\uln}} = 0 ~~,
\eqno(B.3) $$
eliminates the non-covariant terms arising from the variation of the
Chern-Simons term in $~G_{\ula\ulb\ulc}$~ [19].

        We have thus obtained the following superfield equations
$$\li{&R_{\ula\ulb} = e^{-4\Phi/\sqrt3}\, G_{\ula\ulc\uld}
G\du\ulb{\ulc\uld} - {\sqrt3} \eta_{\ula\ulb} \Bo \Phi -
2(\nabla_\ula\Phi)(\nabla_\ulb\Phi)  - 2 e^{-2\Phi/\sqrt3}
F\du\ula{\ulc\, I} F_{\ulb\ulc}{}^I  \cr
& ~~~~~ ~~~~~ + 2i \chi\s_\ulb \nabla_\ula \Tilde\chi + 2i \l^I \s_\ulb
\nabla_\ula\Tilde\l^I + \fracm 2{\sqrt3}
e^{-2\Phi/\sqrt3} ( \chi\s^{\ulb\ulc} \Tilde\l^I +
\Tilde\chi\s^{\ulb\ulc} \l^I ) F_{\ula\ulc}{}^I  \cr
& ~~~~~ ~~~~~ + \fracm i{2\sqrt3} e^{-2\Phi/\sqrt3} \left(
\chi\s^{\ulb\ulc\uld} \Tilde\chi + \l^I \s^{\ulb\ulc\uld} \Tilde\l^I
\,\right)\, G_{\ula\ulc\uld} \cr
& ~~~~~ ~~~~~ + \eta_{\ula\ulb} \bigg[\, -\half e^{-4\Phi /\sqrt3}
(G_{\ulc\uld\ule})^2 + e^{-2\Phi/\sqrt3} (F_{\ulc\uld}{}^I)^2
+(\nabla_\ulc\Phi)^2 \cr
& ~~~~~ ~~~~~ ~~~~~ ~~~~~ - i(\chi\s^\ulc\nabla_\ulc \Tilde\chi) -
i (\l^I\s^\ulc\nabla_\ulc \Tilde\l^I) - \fracm1{\sqrt3} e^{-2\Phi/\sqrt3}
(\chi \s^{\ulc\uld} \l^I + \Tilde\chi\s^{\ulc\uld} \Tilde\l^I )
F_{\ulc\uld}{}^I \cr
& ~~~~~ ~~~~~ ~~~~~ ~~~~~ - \fracm i{4\sqrt3} e^{-2\phi/\sqrt3}
(\chi\s^{\ulc\uld\ule} \Tilde\chi + \l^I \s^{\ulc\uld\ule}
\Tilde\l^I ) G_{\ulc\uld\ule} \,\bigg] ~~,
& (B.4) \cr & ~~~ \cr } $$

\vfill\eject

$$\li{& \fracm i2 (\s^{\ula\ulb\ulc} \Tilde T_{\ulb\ulc} )_\a =
\fracm1{6\sqrt2} e^{-2\Phi/\sqrt3} (\s^{\ulb\ulc\uld} \s^\ula \chi)_\a
G_{\ulb\ulc\uld} - \fracm1{\sqrt2} (\s^\ulb \s^\ula \chi)_\a
\nabla_\ulb \Phi \cr
& ~~~~~ ~~~~~ ~~~~~ ~~~ + \fracm i {2\sqrt2} e^{-\Phi/\sqrt3}
(\s^{\ulb\ulc} \s^\ula \Tilde\l^I)_\a F_{\ulb\ulc}{}^I ~~,
&(B.5a) \cr
& \fracm i2 (\s^{\ula\ulb\ulc} T_{\ulb\ulc} )_{\Dot\a} =
\fracm1{6\sqrt2} e^{-2\Phi/\sqrt3} (\s^{\ulb\ulc\uld} \s^\ula
\Tilde\chi)_{\Dot\a}
G_{\ulb\ulc\uld} - \fracm1{\sqrt2} (\s^\ulb \s^\ula
\Tilde\chi)_{\Dot\a} \nabla_\ulb \Phi \cr
& ~~~~~ ~~~~~ ~~~~~ ~~~ + \fracm i {2\sqrt2} e^{-\Phi/\sqrt3}
(\s^{\ulb\ulc} \s^\ula \l^I)_{\Dot\a} F_{\ulb\ulc}{}^I ~~,
&(B.5b) \cr & ~~~ \cr
& \nabla_\ula (e^{-4\Phi/\sqrt3} G^{\ula\ulb\ulc} ) - \fracm
i{2\sqrt3} \nabla_\ula \left[\, e^{-2\Phi/\sqrt3} (\chi \s^{\ula\ulb\ulc}
\Tilde\chi ) + (\l^I \s^{\ula\ulb\ulc} \Tilde\l^I) \right] ~~,
&(B.6) \cr & ~~~ \cr
& \Bo \Phi +\fracm 1{3\sqrt3} e^{-4\Phi/\sqrt3} (G_{\ula\ulb\ulc})^2 -
\fracm 1{2\sqrt3} (F\du{\ula\ulb}I)^2
- \fracm16 e^{-\Phi/\sqrt3} (\chi\s^{\ula\ulb} \l^I +
\Tilde\chi\s^{\ula\ulb} \Tilde \l^I ) F\du{\ula\ulb}I \cr
& ~~~~~ ~~~~~ - \fracm i{18}
e^{-2\Phi/\sqrt3} ( \chi\s^{\ula\ulb\ulc}\Tilde\chi + \l^I
\s^{\ula\ulb\ulc} \Tilde\l^I ) G_{\ula\ulb\ulc} = 0 ~~,
&(B.7) \cr & ~~~ \cr
& i (\s^\ula\nabla_\ula \Tilde\chi)_\a - \fracm 1{2\sqrt3} e^{-\Phi/\sqrt3}
(\s^{\ula\ulb} \l^I) _\a F\du{\ula\ulb}I - \fracm i{12\sqrt3}
e^{-2\Phi/\sqrt3} (\s^{\ula\ulb\ulc} \Tilde\chi)_\a G_{\ula\ulb\ulc} = 0
{}~~,
&(B.8a) \cr
& i (\s^\ula\nabla_\ula \chi)_{\Dot\a}
- \fracm 1{2\sqrt3} e^{-\Phi/\sqrt3}
(\s^{\ula\ulb} \Tilde\l^I)_{\Dot\a} F\du{\ula\ulb}I - \fracm i{12\sqrt3}
e^{-2\Phi/\sqrt3} (\s^{\ula\ulb\ulc} \chi)_{\Dot\a} G_{\ula\ulb\ulc} = 0 ~~,
&(B.8b) \cr & ~~~ \cr
&\nabla_\ulb (e^{-2\Phi/\sqrt3} F^{\ula\ulb\, I} ) = - i f^{I J K}
(\l^J \s^\ula \Tilde\l^K) - \fracm1{\sqrt3} \nabla_\ulb \left[
e^{-\Phi/\sqrt3} (\chi\s^{\ula\ulb} \l^I +\Tilde\chi\s^{\ula\ulb}
\Tilde\l^I) \right] \cr
& ~~~~~ ~~~~~ ~~~~~ - \fracm 1{12\sqrt3} e^{-4\Phi/\sqrt3}
F\du{\ulb\ulc} I G^{\ula\ulb\ulc} +
\fracm i {72} e^{-2\Phi/\sqrt3} \left[ (\chi\s^{\ula\ulb\ulc}
\Tilde\chi) + ( \l^J \s^{\ula\ulb\ulc} \Tilde\l^J ) \right]
F_{\ulb\ulc}{}^I~~,{~\hskip 0.3in}
&(B.9) \cr & ~~~ \cr
&i(\s^\ula\nabla_\ula\Tilde\l ^I)_\a + \fracm1{2\sqrt3} e^{-\Phi/\sqrt3}
(\s^{\ula\ulb} \chi)_\a F\du{\ula\ulb}I  - \fracm i{12\sqrt3}
(\s^{\ula\ulb\ulc} \Tilde\l^I)_\a e^{-2\Phi/\sqrt3} G_{\ula\ulb\ulc} = 0~~,
&(B.10a) \cr
&i(\s^\ula\nabla_\ula\l^I)_{\Dot\a} + \fracm1{2\sqrt3} e^{-\Phi/\sqrt3}
(\s^{\ula\ulb} \Tilde\chi)_{\Dot\a} F\du{\ula\ulb}I  - \fracm i{12\sqrt3}
(\s^{\ula\ulb\ulc} \l^I)_{\Dot\a} e^{-2\Phi/\sqrt3} G_{\ula\ulb\ulc} =
0~~.
&(B.10b) \cr } $$
These field equations are valid up to trilinear fermionic terms.

        We now take the advantage of these explicit field equations, to see
the consistency of our SD conditions (3.1) - (3.6).
It is now obvious that if we use these SD conditions, all of the above field
equations are consistently satisfied, and yield our previous superfield
equations (3.8) - (3.13).  Even though this statement is confirmed
up to the trilinear fermionic terms in each field equation, we have no reason
to expect any inconsistency when these higher-order fermionic terms are
present, because we have
also a strong support by the satisfaction of all the lower-dimensional
$~(d\le 1)$~ constraints in (2.3).

        There is another lesson we can learn from the above field
equations.  Consider the case of {\it no} SYM multiplet in the
system, and impose {\it only} the condition $~T\du{\ula\ulb}\g = 0$~
in the gravitino equation (B.5).
Then the {\it only} non-trivial
solution for the condition of vanishing spinor source current in (B.5b)
is nothing else than the SDTM (3.5).  Recalling the
fact that there is {\it no} auxiliary field needed for the TM, or in
other words, the TM is by itself {\it off-shell}, we can also conclude
that (3.5) {\it implies} $~\Tilde\chi_{\Dot\a} = 0$~ (3.2).

\bigskip\bigskip

\vfill\eject

\centerline{\bf Appendix C: About Off-Shell Structures}

\medskip\medskip

In this appendix, we first give the {\it off-shell} multiplets of SYM
and SG with appropriate auxiliary fields.  Based on this, we show a
strong evidence that the $~\b\-$functions in our GS string for the SDYM
backgrounds are to vanish, based on a simple superspace argument.

        We start with the {\it off-shell} constraints for the SYM with
{\it global} supersymmetry.  We can expect that the necessary
auxiliary field for this multiplet is to
be a $~D\-$field, according to our experience with the $~D=(1,3),\,N=1$~
case.  Eventually, its form is exactly the same as the {\it purely} SYM
sector in (2.3), with the only exceptional equations:
$$\li{&\nabla_\a \l\du\b I = -\fracm 1{4\sqrt2} (\s^{\ulc\uld})_{\a\b}
\, (F\du{\ulc\uld} I - \half \e\du{\ulc\uld}{\ule\ulf} F\du{\ule\ulf} I )
+ \fracm 1{\sqrt2} C_{\a\b} D^I ~~,
&(C.1a) \cr
&\Tilde\nabla_{\Dot\a}\Tilde\l\du{\Dot\b} I = -\fracm 1{4\sqrt2}
(\s^{\ulc\uld})_{\Dot\a\Dot\b} \,
(F\du{\ulc\uld} I + \half \e\du{\ulc\uld}{\ule\ulf} F\du{\ule\ulf} I )
- \fracm 1{\sqrt2} C_{\Dot\a\Dot\b} D^I ~~,
&(C.1b) \cr
&\nabla_\a D^I = -i(\s^\ula)_{\a\Dot\b} \nabla_\ula \Tilde\l^{\Dot\b\,
I}~~, ~~~~
\Tilde\nabla_{\Dot\a} D^I = + i (\s^\ula)_{\b\Dot\a} \nabla_\ula
\l^{\b\, I} ~~.
&(C.2) \cr }$$

        We now inspect the effect of our SD condition (3.3) on the
$~D\-$auxiliary field.  Since the r.h.s.~of (C.1a) is to vanish due to
$~\l\du\a I=0$~ (3.3). it
automatically follows that
$$D^I = 0 ~~.
\eqno(C.3) $$
This result will be important in the later discussion of finiteness.
For the coupling to SG, we need the SG auxiliary field-dependent
terms in (C.1) and (C.2), as the usual $~D=(1,3)$~ case [20].   However,
for the reason to be made clear
later, we do {\it not} need them for our discussion of finiteness.

        The case of {\it off-shell} SG is also similar to the
$~D=(1,3),\,N=1$~ case [20], and it needs the {\it minimal} auxiliary fields
$~S,~P$~
and $~A_\ula$.  We can easily confirm that the following constraints
satisfy the BIds.:
$$\li{&T\du{\a\Dot\b}\ulc = i(\s^\ulc)_{\a\Dot\b} ~~,
{}~~~~T\du{\ul\a\ulb}\ulc = 0 ~~, ~~~~T\du{\ula\ulb}\ulc = 0~~,
{}~~~~T\du{\ul\a\ul\b}{\ul\g} = 0~~, \cr
&T\du{\a\ulb}\g = + \fracm16 \[ 2\,\d\du\ulb\ulc \,\d\du\a\g + (\s\du\ulb\ulc
)\du\a\g \] A_\ulc ~~, ~~~~
T\du{\Dot\a\ulb}{\Dot\g} = - \fracm16 \[ 2\,\d\du\ulb\ulc
\,\d\du{\Dot\a}{\Dot\g}
+ (\s\du\ulb\ulc )\du{\Dot\a}{\Dot\g} \] A_\ulc ~~, \cr
&T\du{\a\ulb}{\Dot\g} = + \fracm i6 (\s_\ulb)\du\a{\Dot\g} (S+P)~~,
{}~~~~T\du{\Dot\a\ulb}\g = \fracm i6 (\s_\ulb)\du{\Dot\a}\g (S-P)~~,
&(C.4) \cr & ~~~\cr
&\nabla_\a S = +\half (\s^{\ula\ulb} T_{\ula\ulb} )_\a ~~, ~~~~
\Tilde\nabla_{\Dot\a} S = \half (\s^{\ula\ulb} \Tilde T_{\ula\ulb})_{\Dot\a}
{}~~, \cr
&\nabla_\a P = -\half (\s^{\ula\ulb} T_{\ula\ulb} )_\a ~~, ~~~~
\Tilde\nabla_{\Dot\a} P = \half (\s^{\ula\ulb} \Tilde T_{\ula\ulb})_{\Dot\a}
{}~~, \cr
&\nabla_\a A_\ulb = \fracm i 4 (\s\du\ulb{\ulc\uld} \Tilde T_{\ulc\uld}
- 4 \s^\ulc \Tilde T_{\ulb\ulc})_\a ~~,
{}~~~~\Tilde\nabla_{\Dot\a} A_\ulb = - \fracm i 4 (\s\du\ulb{\ulc\uld}
T_{\ulc\uld} - 4 \s^\ulc T_{\ulb\ulc} )_{\Dot\a} ~~,
&(C.5) \cr & ~~~\cr
&R_{\a\b \ulc\uld} = - \fracm 13 (\s_{\ulc\uld})_{\a\b} (S+P) ~~, ~~~~
R_{\Dot\a\Dot\b \ulc\uld} = - \fracm 13 (\s_{\ulc\uld})_{\Dot\a\Dot\b}
(S-P) ~~,\cr
&R_{\a\Dot\b\ulc\uld} = \fracm i3 \e\du{\ulc\uld}{\ule\ulf}
(\s_\ule)_{\a\Dot\b} A_\ulf ~~, \cr
& R_{\a \ulb\ulc\uld} = \fracm i2 (\s_{\[\ulc} \Tilde T_{\uld\] \ulb} -
\s_\ulb \Tilde T_{\ulc\uld} )_\a ~~,
{}~~~~ R_{\Dot\a \ulb\ulc\uld} = \fracm i2 (\s_{\[ \ulc} T_{\uld \] \ulb}
- \s_\ulb T_{\ulc\uld} )_{\Dot\a} ~~.
&(C.6) \cr  } $$
It is also convenient to have the following relations
$$\li{& \Tilde\nabla_{\Dot\a} T\du{\ulb\ulc}\g = - \fracm i6
(\s_{\[\ulb |} )\du{\Dot\a}\g \nabla_{|\ulc\]} (S-P)
+ \fracm i9 (\s_{\[\ulb|} )\du{\Dot\a}\g A_{|\ulc\]}
(S-P) + \fracm i 9 (\s\du{\ulb\ulc}\uld )\du{\Dot\a}\g A_\uld (S-P) \cr
& \nabla_\a T\du{\ulb\ulc}\g = \fracm 13 \d\du\a\g \nabla_{\[\ulb}
A_{\ulc\]}- \fracm 16 (\s\du{\[\ulb|} \uld )\du \a\g \nabla_{|\ulc\]}
A_\uld - \fracm 1{18} (\s_{\ulb\ulc})\du\a\g (S^2- P^2) \cr
& ~~~~~ ~~~~~ ~~~~ + \fracm1{18} \e\du{\ulb\ulc}{\uld\ule}
(\s\du\ule\ulg )\du \a\g A_\ulg A_\ulf - \fracm 14 (\s^{\uld\ule}) \du
\a\g R_{\ulb\ulc\uld\ule} ~~,
&(C.7) \cr & ~~~ \cr
& \nabla_\a \Tilde T\du{\ulb\ulc}{\Dot\g} = - \fracm i6
(\s_{\[\ulb |} )\du\a{\Dot\g} \nabla_{|\ulc\]} (S+P)
- \fracm i9 (\s_{\[\ulb|} )\du\a{\Dot\g} A_{|\ulc\]}
(S+P) - \fracm i 9 (\s\du{\ulb\ulc}\uld )\du\a{\Dot\g} A_\uld (S+P) ~~, \cr
& \Tilde\nabla_{\Dot\a} \Tilde T\du{\ulb\ulc}{\Dot\g} =
- \fracm 13 \d\du{\Dot\a}{\Dot\g} \nabla_{\[\ulb}
A_{\ulc\]}+ \fracm 16 (\s\du{\[\ulb|} \uld )\du{\Dot\a}{\Dot\g}
\nabla_{|\ulc\]} A_\uld - \fracm 1{18} (\s_{\ulb\ulc})
\du{\Dot\a}{\Dot\g} (S^2- P^2) \cr
& ~~~~~ ~~~~~ ~~~~ + \fracm1{18} \e\du{\ulb\ulc}{\uld\ule}
(\s\du\ule\ulg )\du{\Dot\a}{\Dot\g} A_\ulg A_\ulf - \fracm 14 (\s^{\uld\ule})
\du{\Dot\a}{\Dot\g} R_{\ulb\ulc\uld\ule} ~~,
&(C.8) \cr } $$

        As can be easily recognized, there is a parallel structure
between {\it dotted} and {\it undotted} spinors with the
combinations of the auxiliary fields $~S+P$~ and
$~S-P$.  As a universal feature, these two are universally flipped under
the chirality conjugations.

        We can also give the invariant lagrangian for the SG:
$$e \Lag_{\rm SG} = -\fracm14 R - i \left( \psi_\ula\s^{\ula\ulb\ulc}
\nabla_\ulb \Tilde\psi_\ulc\right) - \fracm 16(S^2 - P^2 - A_\ula^2) ~~,
\eqno(C.9) $$
where $~\nabla_{\[\ula} \Tilde\psi\du{\ulb\]} {\Dot\g} = \Tilde
T\du{\ula\ulb} {\Dot\g} \big| - \psi\du{\[\ula|}{\ul\d} T\du{\ul\d |\ulb\]}
{\Dot\g} \big|.\,$\footnotew{See p.~325 of Ref.~[15].}~~Since
the supertranslation structure implied by (C.4)
and (C.5) is similar to the $~D=(1,3)$~ case, our lagrangian (C.9) also
resembles the usual SG lagrangian [20].  However, the opposite sign
between the $~S^2$~ and $~P^2\-$terms reflects our space-time with
the peculiar signature $~(+,+,-,-)$.  This is natural from the following
viewpoints:  First, recall that the $~S~$ and $~P$~ play the same role as
the usual auxiliary fields $~F$~ and $~G$~ in a chiral
multiplet, when supergravity is deleted [20].  Secondly,
in our $~D=(2,2)$~ the kinetic term of chiral superfields is of the
form $~\Phi\Tilde\Phi$, resulting in the {\it mixed} kinetic term
$~\approx A \Bo B$, because $~\Phi$~ and $~\Tilde\Phi$~
are now {\it different} superfields, instead of complex conjugates of each
other.  (See also Refs.~[2,5].)  Therefore in terms of {\it diagonalized}
fields, we get the {\it opposite} signs in two kinetic terms,
accompanied also by the different signs for the $~F^2$~ and
$~G^2\-$terms.
Combined with the first point, this implies the {\it opposite} signs for
the $~S^2$~ and $~P^2\-$terms.

        We now consider what will be the SD condition for the
SG which is consistent with our {\it off-shell} structure.  As such a
condition, we propose
$$R\du{A B \g} \d \equiv \fracm 14 (\s\du \ulc\uld) \du\g\d R_{A B
\uld}{}^\ulc = 0 ~~,
\eqno(C.10) $$
instead of the {\it on-shell} case (3.1).
Eq.~(C.10) is equivalent to the more familiar form $~R_{A B
\ulc\uld} = (1/2) \e\du{\ulc\uld}{\ule\ulf} R_{A B \ule\ulf}$.
In fact, we first see that the $~{\scst A B ~=~ \ul\a\ul\b}$~ case
yields the three conditions
$$ S = - P = \hbox{const}.~, ~~~~ A_\ula = 0 ~~,
\eqno(C.11) $$
while the $~{\scst A B ~=~ \ul\a\ulb}$~ case gives
$$ \li{& T\du{\ula\ulb} \g = 0 ~~,
&(C.12) \cr
&\Tilde T\du{\ula\ulb} {\Dot\g} =
\half \e\du{\ula\ulb}{\ulc\uld} \Tilde T\du{\ulc\uld}{\Dot\g} ~~.
&(C.13) \cr } $$
Eq.~(C.12) coincides with (3.1), while (C.13) implies the SD for the
indices $~{\scst \ula\ulb}$~ of $~\Tilde T\du{\ula\ulb}{\Dot\g}$.
Once these have been established, we easily see that the $~{\scst A B ~=~
\ula\ulb}$~ case is automatically and consistently satisfied, as easily
seen from (C.7) and (C.8).  Notice that the $~{\scst A B~=~\ula\ulb}$~
case is nothing else than the SD condition on the (supercovariant)
Riemann tensor itself.

        It is illustrative to consider the relation
$$i(\s\du\ula{\ulb\ulc} \Tilde T_{\ulb\ulc})_\a = \fracm{4i} 3
(\s_\ula)\du\a{\Dot\b} \Tilde\nabla_{\Dot\b} S + \fracm 43 \nabla_\a
A_\ula~~,
\eqno(C.14) $$
which is obtained from (C.5).  The r.h.s.~is nothing else than the
spinor source current for the gravitino equation.  As we have learned from
(3.8), all such spinor currents are to vanish for the SDSG.  It is also
natural from the fact that the energy-momentum tensor is to vanish, such
that the SD for the Riemann tensor (Ricci-flatness) is maintained, while
the energy-momentum and the spinor current form a current multiplet under
supersymmetry.  From this viewpoint it reasonable to have all the
auxiliary fields to vanish (or constant) as in eq.~(C.11).

        We mention that the TM needs {\it no} auxiliary fields
for its {\it off-shell} formulation.  This is feature is well-known in
the $~D=(1,3),\,N=1$~ case [21], and we can actually confirm this is also the
case with $~D=(2,2),\,N=1$.

        In the context of GS $~\s\-$model, we have to establish the
constraint lagrangian, that gives the SD condition (C.10), as we did in
section 4.  Our proposal is
$$\Lag_{\rm C,\, SG} = \Pi\du\plpl B \Pi\du \mimi A \Pi\du \plpl \d
\Pi\du\mimi \g  R_{A B \g\d} ~~.
\eqno(C.15) $$
Actually this constraint lagrangian yields equations which look
weaker than (C.10).  In fact, by writing each different sorts of indices
in (C.13) explicitly, we see that
the $~{\scst A B ~=~
\a\b,~\Dot\a\b}$~ and $~{\scst A B ~=~
\a\ulb}$~ cases give respectively
$$ R_{\[\a}{}^{\[\b}{}_{\g\]} {}^{\d\]} = 0 ~~,
{}~~~~ R\dud{\Dot\a\b}\b\d = 0 ~~,
{}~~~~ R_{\a\ulb}{}^\a{}_\d = 0 ~~.
\eqno(C.16) $$
However, we easily realize that these conditions give eventually the same
conditions (C.11) - (C.13) as
those from (C.10), even though some indices are contracted and
antisymmetrized.
Thus our total constraint lagrangian in our GS $\s\-$model for the {\it
off-shell} self-dual backgrounds is (4.6) + (4.7) + (C.15).

        Since our analysis of the SG is based on the {\it off-shell}
multiplets, we can conclude
that even the presence of other ``matter'' multiplet will {\it not}
disturb our conditions (C.10).  Therefore, even if we did not include the
$~S,~P$~ and $~A_\ula\-$auxiliary field terms in (C.1) and (C.2) for the SYM,
they do {\it not} do any harm, when talking about the SD conditions of the
SG + SYM.

        We are now ready for the proof of non-corrections to the
$~\b\-$functions for the SYM and SG backgrounds.
There are three main assumptions in our proof.  The first one is
the validity of superspace formulation for the corrections in the SDYM
$~\b\-$functions.  This is empirically true with superstring
corrections, e.g., the $~D=10,\,N=1$~ (heterotic) superstring
corrections [22] at the string tree-level is {\it local} and the superspace
formulation is still valid to accommodate these corrections, and no
counter-example has been known.

        The second assumption is that {\it all} the possible corrections
are embedded into the auxiliary fields of all the multiplets in the off-shell
formulation, and we assume that these {\it minimal} auxiliary fields are
enough to embed
such corrections in the field equations.  This principle is also
analogous to the $~D=10,\,N=1$~ case, where even though no complete
auxiliary fields are known, we could accommodate the possible corrections to
some auxiliary superfields [17,22].  Since the
$~S,~P,~A_\ula$~ and $D\-$field are the {\it complete} auxiliary fields
for the SG and SYM in $~D=(2,2)$, our assumption is
legitimate.\footnotew{In other
words, we have assumed that there will be {\it no} corrections that
modify the dimension ~0~ constraints in the SYM sector.}

        The third assumption is the validity of our constraint
lagrangians (4.6), (4.7) and (C.15) at the quantum level.
This is reasonable, since in the normal-coordinate expansion
after the classical-quantum splittings all the background fields
satisfy the same BIds.~as the starting fields in the GS $~\s\-$model.

        We now see that due to our assumptions, there will be {\it no}
corrections to our auxiliary fields $~S,~P$~ and $~A_\ula$,
implying the {\it absence} of any quantum corrections in the SG sector.  This
in turn implies that the validity of eqs.~(C.1) and (C.2), which results
in the vanishing of $~D^I\-$auxiliary fields, and therefore {\it no}
corrections to the SYM, either.

        There have been various argument about the finiteness of the SDSG and
SDSYM in $~D=(2,2)$~ [5] as well as for the Euclidean case [23].
Our present proof is based
on the {\it off-shell} behaviour
of the multiplets under question, and in this sense it is to be
one of the rigorous proofs for our particular space-time with the
signature $~(+,+,-,-)$.

        The constraint lagrangians we have proposed are to be related to
some vertex operators in the {\it covariant} formulation of the GS
string field theory.  However, since we have {\it not} established such
a covariant
GS string field theory yet, their usefulness will be to give some clue for
inventing some {\it covariant} GS string field
theory.\footnotew{Actually, it is pointed out [24] that
a GS string $~\s-$model can be easily constructed
based on the {\it chiral-superspace} [14], which may be the good
starting point for the GS string field theory.}

\newpage

\centerline{\bf Appendix D: Exact Solution for SDSG + SDSYM + SDTM}

\medskip\medskip

We give in this appendix a set of exact solutions for the coupled system
of SDSG + SDSYM + SDTM in the
self-dual gravitational instanton background by Eguchi and Hanson
(EH) [25].  These solutions have been already presented in Ref.~[10].
Such exact solutions may well play important roles, when we
consider the topological significance of the system, as well as
compactifications into lower-dimensions to get soluble systems [6,7].

        The EH-instanton metric [25] for our SDG is given by
$$\eqalign{d s^2 = \, & {dr^2\over  1 - 1/r^4} - {\frac 14}r^2
\left[  d\vartheta^2 +\sinh^2 \vartheta\,d\f^2\right] \cr
& +{\frac 14}r^2\left( 1 - 1/r^4\right) \left[ d\j^2 + 2\cosh\vartheta\,d\j
d\f + \cosh^2\vartheta\,d\f^2\right]~, \cr}
\eqno(D.1)$$
where our parametrization for the coordinates is
$$z^1 = r\cosh\left( {\frac 12}\vartheta\right) \exp\left[ {\frac 12}i(\j
+ \f) \right]~~,~~~~
z^2 = r\sinh\left( {\frac 12}\vartheta\right) \exp\left[
{\frac 12}i(\j -\f)\right]~~.
\eqno(D.2)$$
In real coordinates we have $~(x^\ulm) = (r,\varphi,\vartheta,\psi)$.

        In our system there are three sets of fermionic fields, the
gravitino $~\Tilde\psi\du\ulm {\Dot\a}$, dilatino $~\chi_\a$~ and gaugino
$~\Tilde\l\du{\Dot\a} I$.  We can simply put all of these fermionic
background fields to zero, for the following reason.
First, We can easily show that the {\it pure-gauge} solution for the
gravitino
$$\Tilde\psi\du\ulm{\Dot\a} = D_\ulm \Tilde\xi^{\Dot\a} ~~,
\eqno(D.3) $$
with an arbitrary space-time dependent spinor $~\Tilde\xi$~ together with
the EH background (D.1) satisfies our field equation (3.1), due to
the identity $~R_{\[\ulm\uln\ulr\]}{}^\uls \equiv 0$.  Secondly, recall
that the supertranslation rule
$$\d\Tilde\psi\du\ulm{\Dot\a} = D_\ulm \Tilde\e^{\Dot\a} ~~, ~~~~
\d e\du\ulm \ula = - i (\e \s^\ula \Tilde\psi_\ulm) ~~,
\eqno(D.4) $$
can completely {\it gauge away} the above solution (D.3).  This means
that an appropriate frame of supersymmetry can put the
background of the gravitino $~\Tilde\psi_\ulm $~ to zero.\footnotew{Of
course, however, this does {\it not} exclude other
gauge-non-trivial solutions.  Our choice corresponds just to
one gauge-trivial family of exact solutions for the gravitino.  We leave
other non-trivial gravitino solutions yet for future studies.}~~Accordingly,
the background loses the manifest supersymmetry, and
other fermions $~\chi_\a$~ and $~\Tilde\l\du \a I$~ are also to put to zero.

        We next solve the SDSYM field equations (3.10) and
(3.13).  There is a convenient method to get a SDYM solution on such a
background, which has been known in the Euclidean case [10].
By this method, we can get a SDYM solution for the $~SU(2)$~
gauge group, which is a subgroup of the
Euclidean Lorentz group $~SO(4) \approx SU(2) \otimes SU(2)$.
In our $~D=(2,2)$~ we choose a gauge group $~SL(2)$, which coincide with
the subgroup of our Lorentz group: $~SO(2,2) \approx SL(2)
\otimes SL(2)$.

        Following this prescription, we identify the Lorentz connection
$~\phi\du\ulm{\ula\ulb}$~ for the EH instanton background (D.1) with our YM
gauge field $~A\du\ulm I$~ as
$$ \eqalign{&\phi\du\ulm{(1)(2)} = \phi\du\ulm {(3)(4)}
\rightarrow A_\ulm{}^1 ~~, \cr
&\phi\du\ulm{(1)(3)} = \phi\du\ulm{(2)(4)} \rightarrow A\du\ulm 2 ~~, \cr
& \phi\du\ulm{(1)(4)} = \phi\du\ulm{(3)(2)}\rightarrow A\du\ulm 3 ~~. \cr }
\eqno(D.5) $$
Here the indices such as $~{\scst (1),~(2),
{}~\cdots}$~ are for the {\it local} Lorentz indices $~{\scst \ula,~\ulb,
{}~\cdots}$, distinguished from the {\it curved} ones $~{\scst
\ulm,~\uln,~\cdots~=~1,~\cdots,~4}$.
This identification has been made possible by
the manifest SD for the $~{\scst \ula\ulb}\-$indices of
$~\phi\du\ulm{\ula\ulb}$.\footnotew{This is {\it not} always true for the
Lorentz connection for any given self-dual Riemann tensor [25].}~~We use
the $~SL(2)$~ gauge group has the generators $~T_I$~ satisfying
$$ \[ T_1,\, T_2\] = - 2T_3~~, ~~~~ \[T_2,\,T_3\] = + 2 T_1~~, ~~~~
\[ T_3,\,T_1\] = -2T_2 ~~.
\eqno(D.6) $$
Now after the identifications (D.5) we can obtain the solution:
$$\eqalign{& A\du 22 = - \half \sqrtrf \sinhth \cos\psi ~~,
{}~~~~A\du 2 3 = - \half \sqrtrf \sinhth \sin \psi ~~,  \cr
&A\du 2 1 = \half \left( 1 + \fracmm1{r^4} \right) \coshth~~, \cr
&A\du 3 2 = \half \sqrtrf \sin\psi ~~, ~~~~
A\du 3 3 = - \half \sqrtrf \cos\psi ~~, \cr
&A\du 4 1 = \half \left( 1 + \fracmm1{r^4} \right) ~~ , \cr }
\eqno(D.7) $$
and all other components are zero.
Since $~SL(2)~$ is non-compact, we always need its metric tensor
$~g{\low{I J}} = \hbox{diag.}\,(1,-1,-1)$~ whenever the indices $~{\scst
I,~J,~\cdots~=~1,~2,~3}$~ are contracted.
The field strength is defined as usual by
$$F_{\ulm\uln}{}^I = \partial_\ulm A\du\uln I - \partial_\uln A\du\ulm
I + f\du{J K} I A\du\ulm J A\du\uln K ~~.
\eqno(D.8) $$
It is easy to confirm the satisfaction of the SD condition (3.13)
explicitly on the EH background
$$\eqalign{&F\du{1 2} 1 = - \fracmm 2{r^5} \coshth ~~, ~~~~
F\du{1 4} 2 = - \fracmm2{r^5} ~~, ~~~~ F\du{23} 1 = - \fracmm 1{r^4}
\sinhth ~~, \cr
&F\du{12} 2 = -\fracmm{\sinhth\cos\psi}{r^5\sqrtrf} ~~,
{}~~~~  F\du{23} 2 = - \fracmm{\sqrtrf \coshth \cos\psi}{2r^4} ~~, \cr
&F\du{34} 2 = \fracmm{\sqrtrf \cos\psi}{2r^4} ~~, ~~~~
F\du{24} 2  = \fracmm{\sqrtrf \sinhth\sin\psi}{2r^4} ~~, ~~~~  \cr
&F\du{12} 3 = - \fracmm{\sinhth \sin\psi}{r^5\sqrtrf}~~, ~~~~
F\du{13} 3  = -\fracmm{\cos\psi}{r^5\sqrtrf} ~~, ~~~~  \cr
&F\du{23} 3 = - \fracmm1{2r^4} \sqrtrf \coshth\sin\psi ~~, ~~~~
F\du{24} 3 = - \fracmm1{2r^4} \sqrtrf \sinhth\cos\psi ~~, ~~~~  \cr
&F\du{34} 3 = \fracmm{\sqrtrf \sin\psi}{2r^4} ~~.   \cr }
\eqno(D.9) $$
All other independent components are zero.
Remarkably, the SD condition (3.6) also holds for
our SDYM instanton, despite of the presence of the EH
gravitational instanton background.
We re-stress the
special role played by the $~SL(2)$~ indices $~{\scst m n}$~ in
$~\phi\du\ulm{\ula\ulb}$~ or $~R\du{\ulm\uln}{\ula\ulb}$, as if they
were ``internal'' $~SL(2)$~ gauge symmetry, that made our prescription
possible.  This is nothing else than
the $~SO(2,2)$~ analog of the usual
Euclidean case $~SO(4) \approx SU(2) \otimes SU(2)$~ [26].

        We finally solve the equations (3.9) and (3.12) for for the
SDTM.  Eq.~(3.9) is already satisfied by the fermionic trivial
solutions.  We rewrite (3.12) as
$$\Bo \phi - \fracm 13 F\du{\ulm\uln} I F\ud{\ulm\uln} I = 0 ~~,
\eqno(D.10) $$
in terms of
$$~\phi \equiv \exp\left(\fracm2{\sqrt3}\Phi\right)~~.
\eqno(D.11) $$
Inserting the solution (D.10) in (D.12), we get
$$\phi''(r) + \fracmm{3r^4+1}{r(r^4-1)} \phi'(r) =
\fracmm{32}{r^8(r^4-1)} ~~.
\eqno(D.12) $$
Here a prime is for each derivative of $~d/dr$.  Eq.~(D.14) is
easily solved by
$$\li{&\phi = - \fracmm{2(3r^4+1)}{3r^6} + \fracmm{a - 4}4 \ln\left(
\fracmm{r^2+1}{r^2-1}\right) + 1 ~~,
&(D.13) \cr
&\Phi = \fracm{\sqrt3}2 \ln\left[ - \fracmm{2(3r^4+1)}{3r^6}
+ \fracmm{a-4}4 \ln\left( \fracmm{r^2 - 1}{r^2 + 1} \right) + 1 \right]
{}~~,
&(D.14) \cr } $$
under the boundary condition $~\Phi (r\rightarrow \infty) = 0$.
The $~a$~ is an arbitrary integration constant.  Now the $~B_{\ulm\uln}$~
field is solved {\it via} its field strength as
$$ \li{&G_{2 3 4} =  \fracmm {{\sqrt3}(4-ar^8)}{16r^8} \sinhth~~,
&(D.15) \cr
&B_{2 4} = \fracmm{4 + 3a}{16\sqrt3} \coshth + B_0 ~~,
&(D.16)  \cr } $$
with an arbitrary constant $~B_0$.  Other independent components of
$~G_{\ulm\uln\ulr}$~ and $~B_{\ulm\uln}$~ are zero.
These solutions are also consistently with the Bianchi identity:
$$\partial_{\[ \ulm} G_{\uln\ulr\uls\]} = - \fracm{\sqrt3}2
F\du{\[\ulm\uln} I F_{\ulr\uls\]\, I} ~~.
\eqno(D.17) $$

        The prescription of getting the $~SL(2)$~ SDSYM field
from the SDG solution is universal for other cases,
e.g., the Taub-Nut solution [27].  However, in some cases,
the Lorentz connection $~\phi\du\ulm{\ula\ulb}$~ is {\it not manifestly}
self-dual, even though $~R\du{\ulm\uln}{\ula\ulb}$~ {\it is}
self-dual.  In such cases, we have to arrange $~\phi\du\ulm{\ula\ulb}$~ by
appropriate Lorentz transformation such that
the new Lorentz connection is manifestly self-dual [25].  After such a
redefinition of $~\phi$'s, the identification (D.5) is straightforward to get
an exact solution for $~SL(2)~$ SDSYM.

        A remarkable point about these solutions is their compatibility
with our $~N=2$~ superstring.
Our Wess-Zumino-Witten (WZW) term in (4.2) is\footnotew{Compared with
(4.2), we need the standard overall factor $~1/(2\pi)$~ [28] in
accordance with the topological background.}
$$ I_{\rm WZW} = \int d^2 \s\, \left[- \fracm1{2\pi}\fracm 1{\sqrt3} \,
\e^{i j} \Pi\du i A \Pi \du j B B_{B A}\, \right] ~~.
\eqno(D.18) $$
In the absence of fermionic backgrounds
we can replace $~\Pi \du i A$~ by $~(\partial_i X^\m) e\du\m m$~
(Neveu-Ramond-Schwarz $~\s\-$model).
To see the effect of our exact solutions on the WZW-term,
we regard the $~r\-$coordinate as a ``time'' coordinate for our
``instanton'' solution (D.9).  The
effect of such ``instanton'' at time $~r$~ results in the exponent
in the path-integral \footnotew{Notice that the effect of YM gauge
anomaly is included in the CS
term by the GS mechanism [28], when $~3\partial_{\[\m} B_{\n\r\]}$~ is
converted into $~\Hat G_{\m\n\r}$~ in (D.19).  See Ref.~[24] for the
details.}
$$\eqalign{ P(r) &= \fracm1{2\pi}\fracm1{\sqrt3} \int d^2 \s \, \e^{i j}
(\partial_i X^\m) (\partial_j X^\n) B_{\m\n} \cr
& = \fracm1{2\pi} \fracm 1{\sqrt3} \int d^2\s \int_0^1 du \, \e^{\hat
i\hat j\hat k} (\partial_{\hat i} \Hat X^\m) (\partial_{\hat j} \Hat X^\n)
(\partial_{\hat k} \Hat X^\r) \Hat G_{\m\n\r} \cr
& = \fracmm{3(4-ar^8)} {16\pi r^8} \int _0^{2\pi} d\varphi \int_0^{2\pi}
d\psi \int_0^\pi d\theta \sin\theta \cr
& = \fracmm{6\pi}{r^8} - \fracmm{3a\pi}2 ~~, \cr}
\eqno(D.19) $$
in the string path-integral [29].  We have introduced
a new third coordinate $~0\le u\le 1$~ in addition to the
two-dimensional world-sheet coordinates $\,(\s^i)$, following what is
called ``Vainberg construction'' [22,29].  Accordingly all the quantities
with {\it hats} are in total extended three-dimensional
manifold.  In particular, $~\Hat G_{\m\n\r}$~ is a function of
$~\Hat X^\m(\s,u)$, satisfying $~\Hat X^\m(\s,0) = 0,~\Hat X^\m(\s,1)
= X^\m(\s)$~ [22,29].
In the third equality in (D.19), we have also identified these
three-dimensional coordinates
with the $~D=(2,2)$~ ones by $~\s^1 = \varphi,~ \s^2 = \theta,~
u = \psi /(2\pi)$.  Then the integrand contains nothing else than the
Jacobian of this ``coordinate transformation'' from
$~(\s^1,\,\s^2,\,u)$~ to $~(\Hat X^\m)$.
Eventually the total ``phase-shift'' in the string path-integral
by our in the ``time'' interval $~1\le r <\infty$~ is [29]
$$P(\infty) - P(1) = - 6\pi~~.
\eqno(D.20) $$
After a Wick-rotation, this amounts to be a multiple of $~2\pi i$, as
desired.  This implies the consistency of
exact solutions with the $~N=2$~ superstring as its background {\it via}
the WZW-term!  This result has strong topological
significance and consistency of the $~N=2$~ string theory living on
our peculiar SDSG + SDSYM + SDTM background.

\newpage

\refs

\normalsize

\items{1} H. Nishino, S. J. Gates, Jr.,
Maryland preprint, \mpl{7}{92}{2543}.

\items{2} S.~V.~Ketov, S.~J.~Gates, Jr.~and H.~Nishino, Maryland
preprint, UMDEPP 92--163 (February 1992).

\items{3} H. Nishino, S. J. Gates, Jr. and S. V. Ketov,
Maryland preprint, UMDEPP 92--171 (February 1992).

\items{4} S. J. Gates, Jr., H. Nishino and S. V. Ketov, Maryland
preprint, UMDEPP 92--187 (March 1992), to appear in Phys.~Lett.~B.

\items{5} S.V.~Ketov, H.~Nishino and S.J.~Gates, Jr., Maryland preprint,
UMDEPP 92--211 (June, 1992), to appear in Nucl.~Phys.~B.

\items{6} M. F. Atiyah, unpublished;
\item{  } R. S. Ward, Phil.~Trans.~Roy.~Lond.~{\bf A315} (1985) 451;
\item{  } N. J. Hitchin, Proc.~Lond.~Math.~Soc.~{\bf 55} (1987) 59.

\items{7} A.A.~Belavin, A. M. Polyakov, A. Schwartz and Y. Tyupkin,
\pl{59}{75}{85};
\item{  } R.S.~Ward, \pl{61}{77}{81};
\item{  } M.F.~Atiyah and R.S.~Ward, \cmp{55}{77}{117};
\item{  } E.F.~Corrigan, D.B.~Fairlie, R.C.~Yates and P.~Goddard,
\cmp{58}{78}{223};
\item{  } E.~Witten, \prl{38}{77}{121};
\item{  } A.N.~Leznov and M.V.~Saveliev, \cmp{74}{80}{111};
\item{  } L.~Mason and G.~Sparling, \pl{137}{89}{29};
\item{  } I.~Bakas and D.A.~Depireux, \mpl{6}{91}{399}; {\it ibid.} 1561;
2351.

\items{8} H. Ooguri and C. Vafa, \mpl{5}{90}{1389};
\np{361}{91}{469}; \ibid{B367}{91}{83}.

\items{9} S.J.~Gates and H.~Nishino, Maryland preprint, UMDEPP 93--51
(Oct.~1992).

\items{10} H.~Nishino, Maryland preprint, UMDEPP 93--52 (Sept.~1992).

\items{11} M. T. Grisaru, H. Nishino and D. Zanon, \pl{206}{88}{625};
\np{314}{89}{363}.

\items{12} H. Nishino, \np{338}{90}{386}.

\items{13} W.~Siegel, Stony Brook preprint, ITP-SB-92-31 (July 1992);
and private communications.

\items{14} A.~Parkes, Z{\" u}rich preprint, ETH-TH/92-14 (March, 1992);
\item{  } W.~Siegel, Stony Brook preprint, ITP-SB-92-24 (May, 1992).

\items{15} S.J. Gates Jr., M.T. Grisaru, M. Rocek and W. Siegel, {\it
Superspace}, Benjamin/Cummings, Reading, MA, 1983.

\items{16} H. Nishino, \pl{214}{88}{374}.

\items{17} S.J.~Gates, Jr.~and S.~Vashakidze, \np{291}{87}{172}.

\items{18} T.~Kugo and P.~Townsend, \np{221}{83}{357}.

\items{19} H.~Nishino and E.~Sezgin, \np{278}{86}{353}.

\items{20} P.~van Nieuwenhuizen, \prep{68}{81}{189};
\item{  } A.~Salam and E.~Sezgin, {\it ``Supergravities in Diverse
Dimensions''}, Elsevier Science
Publishings, B.~V.~and World Scientific Pub.~Co.~Pte.~Ltd.~(1989).

\items{21} W.~Siegel, \pl{85}{79}{333};
S.J.~Gates, Jr., \np{184}{81}{381}.

\items{22} S.J.~Gates, Jr.~and H.~Nishino, \pl{173}{86}{46} and 52;
\np{291}{87}{205};
\item{   } H.~Nishino and S.J.~Gates, \np{282}{87}{1}.

\items{23} C.M.~Christensen, S.~Deser, M.J.~Duff and M.T.~Grisaru,
\pl{84}{79}{411}; R.E.~Kallosh, JETP Lett.~{\bf 29} (1979) 172;
\np{165}{80}{119}.

\items{24} W.~Siegel, Stony Brook preprint, ITP-SB-92-53 (October,
1992).

\items{25} T.~Eguchi and A.~Hanson, \ap{120}{79}{82};
E.~Calabi, Ann.~Sci.~Ec.~Norm.~Sup.~{\bf 12} (1979) 269;
T.~Eguchi, P.B.~Gilkey and A.J.~Hanson, \prep{66}{80}{213}.

\items{26} J.M.~Charap and M.~Duff, \pl{69}{77}{445}.

\items{27} S.W.~Hawking, Phys.~Lett.~{\bf 60A} (1977) 81.

\items{28} See e.g., M.B.~Green, J.H.~Schwarz and E. Witten, {\it ``Superstring
Theory''}, Vols.~I and II, Cambridge University Press (1987).

\items{29} R.~Rohm and E.~Witten, \ap{170}{86}{454}.

\end{document}